\documentclass[oribibl]{llncs}
\date{}
\usepackage{multicol}
\usepackage{amsmath}
\usepackage{amssymb}
\usepackage{graphicx}
\usepackage{epsfig}

\newcommand{\boxtheorem}{\hfill $\Box$}
\newcommand{\nit}[1]{{\it #1}}

\newcommand{\IC}{\nit{IC}}

\def\iR{{\cal R }}
\def\iD{{\it D }}
\def\iA{{\mathcal A }}
\def\IC{{\textit{IC}}}
\newcounter{lemmaA-counter}
\newcounter{propositionA-counter}
\newenvironment{lemmaA}%
{\vskip \abovedisplayskip \refstepcounter{lemmaA-counter}%
\noindent {\bf Lemma A.\arabic{lemmaA-counter}.}}%

\newenvironment{propositionA}%
{\vskip \abovedisplayskip \refstepcounter{propositionA-counter}%
\noindent {\bf Proposition A.\arabic{propositionA-counter}.}}%

\newcommand{\defproof}[2]{{\noindent\bf Proof of #1:\
}#2 \boxtheorem}

\newcommand{\dproof}[2]{{\noindent\bf Proof:\
}#2 \boxtheorem}

\title{\vspace*{-1cm}
\bf Complexity and Approximation of Fixing Numerical Attributes in
Databases Under Integrity Constraints\thanks{Dedicated to the
memory of Alberto Mendelzon. Our research on this topic started
with conversations between Loreto Bravo and him. Alberto was
always generous with his time, advice and ideas; our community is
already missing him very much.}}
\author{{\bf Leopoldo Bertossi},~~
 {\bf Loreto Bravo}\\
 Carleton University,
School of Computer Science, Ottawa, Canada. \newline
\hspace*{1.5cm}\{bertossi,lbravo\}@scs.carleton.ca \newline {\bf
Enrico Franconi},~~ {\bf Andrei Lopatenko}\thanks{Also: University
of Manchester, Department of Computer Science, UK.}\\Free
University of Bozen--Bolzano,
Faculty of Computer Science, Italy.\\
\{franconi,lopatenko\}@inf.unibz.it}

\institute{}

\begin{document}
\pagestyle{empty} 
\maketitle
\thispagestyle{empty}

\vspace{-7mm}\begin{abstract} Consistent query answering is the
problem of computing the answers from a database that are
consistent with respect to certain integrity constraints that the
database as a whole may fail to satisfy. Those answers are
characterized as those that are invariant
  under minimal forms of restoring the consistency of the database. In this context, we study the
problem of repairing databases by fixing integer numerical values
at the attribute level with respect to denial and aggregation
constraints. We introduce a quantitative definition of database
fix, and investigate the complexity of several decision and
optimization problems, including  {\em DFP}, i.e. the existence of
fixes within a given distance from the original instance, and {\em
CQA}, i.e. deciding consistency of answers to aggregate
conjunctive queries under different semantics. We provide sharp
complexity bounds, identify relevant tractable cases; and
introduce approximation algorithms for some of those that are
intractable. More specifically, we obtain results like
undecidability of existence of fixes for aggregation constraints;
{\em MAXSNP}-hardness of {\em DFP}, but a good approximation
algorithm for a relevant special case; and intractability but good
approximation for {\em CQA} for aggregate queries for one database
atom denials (plus built-ins).
\end{abstract}

\vspace*{-6mm}
\section{Introduction}\label{sec:intro}
\vspace*{-2mm} Integrity constraints (ICs) are used to impose
semantics on a database with the purpose of making the database an
accurate model of an application domain. Database management
systems or application programs enforce the satisfaction of the
ICs by rejecting undesirable updates or executing additional
compensating actions. However, there are many situations where we
need to interact with databases that are inconsistent in the sense
that they do not satisfy certain desirable ICs. In this context,
an important problem in database research consists in
characterizing and retrieving consistent data from inconsistent
databases \cite{bookChapter}, in particular consistent answers to
queries. From the logical point of view, consistently answering a
query posed to an inconsistent database amounts to evaluating the
truth of a formula against a particular {\em class of first-order
structures} \cite{ABC99}, as opposed to the usual process of truth
evaluation in a single structure (the relational database).

Certain database applications, like census, demographic,
financial, and experimental data, contain quantitative data,
usually associated to nominal or qualitative data, e.g. number of
children associated to a household identification code (or
address); or measurements associated to a sample identification
code. Usually this kind of data contains errors or mistakes with
respect to certain semantic constraints. For example, a census
form for a particular household may be considered incorrect if the
number of children exceeds 20; or if the age of a parent is less
than 10. These restrictions can be expressed with denial integrity
constraints, that prevent some attributes from taking certain
values \cite{franconi}. Other restrictions may be expressed with
aggregation ICs, e.g. the maximum concentration of certain toxin
in a sample may not exceed a certain specified amount; or the
number of married men and married women must be the same.
 Inconsistencies in numerical data can be
resolved by changing individual attribute values, while
 keeping values in the keys, e.g. without changing the household
code, the number of children is decreased considering the
admissible values.

We consider the problem of fixing integer numerical data wrt
certain constraints while (a) keeping the values for the
attributes in the keys of the relations, and (b) minimizing the
quantitative global distance between the original and modified
instances. Since the problem may admit several global solutions,
each of them involving possibly many individual changes, we are
interested in characterizing and computing data and properties
that remain invariant under any of these fixing processes. We
concentrate on
 denial and aggregation constraints;
and conjunctive queries, with or without aggregation.

Database repairs have been  studied in the context of consistent
query answering (CQA), i.e. the process of obtaining the answers
to a query that are consistent wrt  a given set of ICs
\cite{ABC99} (c.f. \cite{bookChapter} for a survey). There,
consistent data is characterized as invariant under all minimal
forms of restoring consistency, i.e. as data that is present in
all minimally repaired versions of the original instance (the {\em
repairs}). Thus, an answer to a query is consistent if it can be
obtained as a standard answer to the query from {\em every
possible} repair. In most of the research on CQA, a repair
 is a new instance that satisfies
the given ICs, but differs from the original instance by a minimal
set, under set inclusion, of (completely) deleted or inserted
tuples. Changing the value of a particular attribute can be
modelled as a deletion followed by an insertion, but this may not
correspond to a minimal repair. However, in certain applications
it may make more sense to correct (update) numerical values only
in certain attributes. This requires a new definition of repair
that considers: (a) the quantitative nature of individual changes,
(b) the association of the numerical values to other key values;
and (c) a quantitative distance between database instances.

\begin{example}\label{ex:net}
Consider a network traffic database $D$ that stores flow
measurements
 of links in a network. This network has two types of
links, labelled $0$ and $1$, with maximum capacities $1000$ and
$1500$, resp.

\begin{tabular}{c|c|c|c|c|}
\hline \multicolumn{1}{|c|}{Traffic}&Time&Link& Type &Flow~\\
\hline &1.1 & a & 0 &1100 \\
\cline{2-5} &1.1& b & 1 & 900\\
\cline{2-5} &1.3& b & 1 & 850\\\cline{2-5}
\end{tabular}

\noindent Database $D$ is inconsistent wrt this IC. Under the
tuple and set oriented semantics of repairs \cite{ABC99}, there is
a unique repair, namely deleting tuple $\nit{Traffic}(1.1,a,0,$
$1100)$. However, we have two options that may make more sense
than deleting the flow measurement, namely updating the violating
tuple to $\nit{Traffic}(1.1,a, 0,$ $1000)$ or to
$\nit{Traffic}(1.1,a, 1,1100)$; satisfying an implicit requirement
that the numbers should not change too much.\boxtheorem
\end{example}

\vspace{-2mm}\noindent Update-based repairs for restoring
consistency are studied in \cite{Wij03}; where changing values in
attributes in a tuple is made a primitive repair action; and
semantic and computational problems around CQA are analyzed from
this perspective. However, peculiarities of changing numerical
attributes are not considered, and more importantly, the distance
between databases instances used in \cite{Wij03,Wij04} is based on
set-theoretic homomorphisms, but not quantitative, as in this
paper. In \cite{Wij03} the repaired instances  are called {\em
fixes}, a term that we keep here (instead of {\em repairs}),
because our basic repair actions are also changes of (numerical)
attribute values. In this paper we consider fixable attributes
that take integer values and the quadratic, Euclidean distance
$L_2$ between database instances.  Specific fixes and
approximations may be different under other distance functions,
e.g. the ``city distance" $L_1$ (the sum of absolute differences),
 but the general (in)tractability and approximation results
remain. However, moving to the case of real numbers will certainly
bring new issues that require different approaches; they are left
for ongoing and future research. Actually it would be natural to
investigate them in the richer context of constraint databases
\cite{kuper}.

The problem of attribute-based correction of census data forms
 is
addressed in \cite{franconi} using disjunctive logic programs with
stable model semantics. Several underlying and implicit
assumptions that are necessary for that approach to work are made
explicit and used here, extending the semantic framework of
\cite{franconi}.

We provide semantic foundations for fixes that are based on
changes on numerical attributes in the presence of key
dependencies and wrt denial and aggregate ICs, while keeping the
numerical distance to the original database to a minimum. This
framework introduces new challenging decision and optimization
problems, and many algorithmic and complexity theoretic issues. We
concentrate in particular on the ``Database Fix Problem" ({\em
DFP}), of determining the existence of a fix at a distance not
bigger than a given bound, in particular considering the problems
of construction and verification of such a fix. These problems are
highly relevant for large inconsistent databases. For example,
solving {\em DFP} can help us find the minimum distance from a fix
to the original instance; information that can be used to prune
impossible branches in the process of materialization of a fix.
The {\em CQA} problem of deciding the consistency of query answers
is studied wrt decidability, complexity, and approximation under
several alternative semantics.

We prove that {\em DFP} and {\em CQA}   become undecidable in the
presence of aggregation constraints. However,  {\em DFP} is {\em
NP}-complete for linear denials, which are enough to capture
census like applications. {\em CQA} belongs to $\Pi^P_2$ and
becomes $\Delta^P_2$-hard, but for a relevant class of denials  we
get tractability of CQA to non aggregate queries, which is again
lost with aggregate queries. Wrt
 approximation, we prove that {\em DFP} is $\nit{MAXSNP}$-hard
 in general, and for a relevant subclass of denials we provide an
approximation within a constant factor that depends on the number
of atoms in them. All the algorithmic and complexity results,
unless otherwise stated, refer to data complexity \cite{AHV95},
i.e. to the size of the database that here includes a binary
representation for numbers.  For complexity theoretic definitions
and classical results we refer to \cite{papadimitriou94}.

This paper is structured as follows. Section \ref{sec:prel}
introduces basic definitions. Sections \ref{sec:MinDelta} presents
the notion of database fix, several  notions of consistent answer
to a query; and some relevant decision problems. Section
\ref{sec:Complexity} investigates their complexity. In Section
\ref{sec:approx} approximations for the problem of finding the
minimum distance to a fix are studied, obtaining negative results
for the general case, but good approximation for the class of
local denial constraints. Section \ref{sec:tractable} investigates
tractability of {\em CQA} for conjunctive queries and denial
constraints containing one database atom plus built-ins. Section
\ref{sec:conclusions} presents some conclusions and refers to
related work. Proofs and other auxiliary, technical results can be
found in  Appendix \ref{sec:proof}.

\vspace*{-3mm}\section{Preliminaries}\label{sec:prel}
\vspace{-2mm} Consider a relational schema $\Sigma=({\cal U},
{\cal R}, {\cal B}, {\cal A})$, with domain ${\cal U}$ that
includes $\mathbb{Z},$\footnote{With simple denial constraints,
numbers can be restricted to, e.g. $\mathbb{N}$ or  $\{0,1\}$.}
${\cal R}$ a set of database predicates, ${\cal B}$ a set of
built-in predicates, and ${\cal A}$ a set of attributes. A
database instance is a finite collection $D$ of {\em database
tuples}, i.e. of ground atoms $P(\bar{c})$, with $P \in \iR$ and
$\bar{c}$ a tuple of constants in $\cal U$. There is a set ${\cal
F} \subseteq \iA$ of all the {\em fixable} attributes, those that
take values in $\mathbb{Z}$ and are allowed to be fixed.
Attributes outside ${\cal F}$ are called {\em rigid}. ${\cal F}$
need not contain all the numerical attributes, that is we may also
have rigid numerical attributes.

We also have a set ${\cal K}$ of key constraints  expressing that
 relations  $R \in \iR$ have a primary key $K_{\!R}$, $K_{\!R}
\subseteq (\iA \smallsetminus {\cal F})$. Later on (c.f.
Definition \ref{def:LSrep}), we will assume that ${\cal K}$ is
satisfied both by the initial instance $D$, denoted $D \models
{\cal K}$, and its fixes. Since ${\cal F} \cap K_{\!R} =
\emptyset$, values in key attributes cannot be changed in a fixing
process; so the constraints in $\cal K$ are {\em hard}. In
addition, there may be a separate set $\IC$ of {\em flexible} ICs
that may be violated, and it is the job of a fix to restore
consistency wrt them (while still satisfying $\cal K$).

A \emph{linear denial constraint} \cite{kuper} has  the form
~$\forall \bar{x} \neg(A_1\wedge \ldots\wedge A_m)$, where the
$A_i$ are database atoms (i.e. with predicate in $\iR$), or
built-in atoms of the form  $x \theta c$, where $x$ is a variable,
$c$ is a constant and $\theta \in \{=,$ $\neq$, $<$, $>,$ $\leq,$
$\geq\}$, or $x=y$. If $x\neq y$ is allowed, we call them {\em
extended} linear denials.

\vspace{-2mm}
\begin{example}\label{ex:linear} The following are
linear denials (we replace $\wedge$ by a comma): (a)
 No customer is younger than 21:~ $\forall \nit{Id},$ $\nit{Age,Income,Status} \neg(\nit{Customer(Id,
 Age},$
$\nit{Income},$ $\nit{Status}), \nit{Age} < 21 )$. (b) No customer
with income less than 60000 has ``silver" status: $\forall
\nit{Id, Age,}$ $\nit{Income, Status} \neg(\nit{Customer(Id,
Age},$ $\nit{Income,
 Status)},$ $\nit{Income}$ $ < 60000,  \nit{Status} =
 \nit{silver})$. (c) The constraints in Example \ref{ex:net},
 e.g.\\
 $\forall \nit{T,L,Type,Flow} \neg(\nit{Traffic}(T,L,$ $\nit{Type,Flow)},$ $ \nit{Type}=0, $
 $\nit{Flow}>1000)$.
 \boxtheorem
\end{example}
We consider   aggregation constraints (ACs) \cite{rsss-98} and
aggregate queries with {\em sum, count, average}. {\em Filtering}
ACs impose conditions on the tuples over which aggregation is
applied, e.g. $\nit{sum}(A_1\!: A_2=3)\
>\ 5$ is a sum  over $A_1$ of tuples with
$A_2 =3$. {\em Multi-attribute} ACs allow arithmetical
combinations of attributes as arguments for $\nit{sum}$, e.g.
$sum(A_1 + A_2)> 5$ and $sum(A_1 \times A_2) > 100$. If an AC has
attributes from more than one relation, it is {\em
multi-relation}, e.g. $sum_{R_1}(A_1) = sum_{R_2}(A_1)$, otherwise
it is {\em single-relation}.

An {\em aggregate conjunctive query} has the form $q(x_1, \ldots
x_m;~\nit{agg}(z)) \leftarrow B(x_1,$ $\ldots,$ $x_m,$ $z,y_1,
\ldots,$ $ y_n)$, where $\nit{agg}$ is an aggregation function and
its {\em non-aggregate matrix} (NAM) given by $q^\prime(x_1,
\ldots x_m) \leftarrow B(x_1, \ldots, x_m,$ $z$, $y_1, \ldots,$ $
y_n)$ is a usual first-order (FO) conjunctive query with built-in
atoms, such that the {\em aggregation attribute} $z$ does not
appear among the $x_i$. Here we use the set semantics. An
aggregate conjunctive query is {\em cyclic} ({\em acyclic}) if its
NAM is cyclic (acyclic)  \cite{AHV95}.

\begin{example}\label{ex:acq} ~$q(x,y,sum(z))
\leftarrow R(x,y), Q(y,z,w),$ $w \neq 3$ is an aggregate
conjunctive query, with aggregation attribute $z$. Each answer
$(x,y)$ to its NAM, i.e. to $q(x,y) \leftarrow R(x,y), Q(y,z,w), w
\neq 3$, is expanded to $(x,y,sum(z))$ as an answer to the
aggregate query. ~$sum(z)$ is the sum of all the values for $z$
having a $w$, such that $(x,y,z,w)$ makes $R(x,y), Q(y,z,w), w
\neq 3$ true. In the database instance $D = \{R(1,2),$ $R(2,3),$ $
Q(2,5,9), Q(2,6,7),$ $ Q(3,1,1), Q(3,1,5),$ $Q(3,8,3)\}$ the
answer set for the aggregate query is $\{(1,2,5+6), (2,3,1+1)\}$.
\boxtheorem
\end{example}

\vspace{-2mm}\noindent An {\em aggregate comparison query} is a
sentence of the form $q(\nit{agg}(z)), \nit{agg}(z) \theta k$,
where $q(\nit{agg(z)})$ is the head of a scalar aggregate
conjunctive query (with no free variables), $\theta$ is a
comparison operator, and $k$ is an integer number. For example,
the following is an aggregate comparison query asking whether the
aggregated value obtained via $q(sum(z))$ is bigger than 5: $Q\!:
q(sum(z)), sum(z)
> 5$, with $q(sum(z)) \leftarrow R(x,y), Q(y,z,w),w \neq 3$.

\vspace{-4mm}
\section{Least Squares Fixes}\label{sec:MinDelta}

\vspace{-2mm} When we update numerical values to restore
consistency, it is desirable to make the smallest overall
variation of the original values, while considering the relative
relevance or specific scale of each of the fixable attributes.
Since the original instance and a fix will share the same key
values (c.f. Definition \ref{def:LSrep}), we can use them to
compute variations in the numerical values. For a tuple $\bar{k}$
of values for the key $K_{\!R}$ of relation $R$ in an instance
$D$, $\bar{t}(\bar{k},R,D)$ denotes the unique tuple $\bar{t}$ in
relation $R$ in instance $D$ whose key value is $\bar{k}$. To each
attribute $A \in {\cal F}$ a fixed numerical weight
$\alpha_{\!_A}$ is assigned.

\vspace{-2mm}
\begin{definition}  \label{def:dist} \em For instances
$\iD$ and $\iD^\prime$ over schema $\Sigma$ with  the same set
$\nit{val}(K_{\!R})$ of tuples of key values
 for each relation $R \in {\cal R}$, their \textit{square distance}
 is\\ \hspace*{1.5cm}$\Delta_{\bar{\alpha}}(\iD, \iD^\prime) =  \sum_{\stackrel{R \in \iR,A \in {\cal F}}{\bar{k}
 \in \nit{val}(K_{\!R})}}\alpha_{\!_A}
    (\pi_{\!_A}(\bar{t}(\bar{k},R,D))-
    \pi_{\!_A}(\bar{t}(\bar{k},R,D^\prime)))^2,$\\
 where $\pi_{\!_A}$ is the projection on attribute $A$ and
$\bar{\alpha} = (\alpha_{\!_A})_{A \in {\cal F}}$. \boxtheorem
\end{definition}

\begin{definition}\label{def:LSrep} \em For an instance $D$, a
set of fixable attributes ${\cal F}$, a set of key dependencies
${\cal K}$, such that $D \models {\cal K}$, and a set of flexible
ICs $\IC$: ~A {\em fix}  for $D$ wrt $\IC$ is an instance
$D^\prime$ such that: (a) $D^\prime$ has the same schema and
domain as $D$; (b) $D^\prime$ has the same values as $D$ in the
attributes in $\iA \smallsetminus {\cal F}$; (c) $D^\prime \models
{\cal K}$; and (d) $D^\prime \models \IC$.~ A {\em least squares
fix} (LS-fix) for $D$ is a fix $D^\prime$ that minimizes the
square distance $\Delta_{\bar{\alpha}}(D,D^\prime)$ over all the
instances that satisfy (a) - (d).
 \boxtheorem
\end{definition}
In general we are interested in LS-fixes, but (non-necessarily
minimal) fixes will be useful auxiliary instances.

\vspace{-2mm} \begin{example}\label{ex:net2} (example \ref{ex:net}
cont.) $\iR =$ $\{\nit{Traffic}\},$ $\iA = \{Time,$ $Link,$
$Type,$ $ Flow\}$, $K_{\!\nit{Traffic}} = \{Time, Link\}$, ${\cal
F} = $ $\{Type, $ $Flow\}$, with weights ${\bar{\alpha}}=
(10^{-5}, 1)$, resp. For original instance $D$,
$\nit{val}(K_{\!\nit{Traffic}})$ $= \{(1.1,a),(1.1,b),(1.3,b)\},$
$\bar{t}((1.1,a),$ $\nit{Traffic},D) = (1.1,a,0,1100)$, etc. Fixes
are $D_1=\{(1.1,a,0,1000),$ $(1.1,b,1,900),$ $ (1.3,b,1,850)\}$
and $D_2=\{(1.1,a,1,1100),$ $(1.1,b,1,900),$ $ (1.3,b,1,850)\}$,
with distances ~$\Delta_{\bar{\alpha}}(D,D_1) $ $=100^2 \times
10^{-5}= 10^{-1}$ and ~$\Delta_{\bar{\alpha}}(D,D_2)= 1^2 \times
1$, resp. Therefore, $D_1$ is the only LS-fix. \boxtheorem
\end{example}
\vspace{-1mm} The coefficients $\alpha_{\!_A}$ can be chosen in
many different ways depending on factors like relative relevance
of attributes, actual distribution of data, measurement scales,
etc.  In the rest of this paper we will assume, for
simplification, that $\alpha_{\!_A}=1$ for all $A \in {\cal F}$
and $\Delta_{\bar{\alpha}}(\iD, \iD^\prime)$ will be simply
denoted by $\Delta(\iD, \iD^\prime)$.

\vspace{-1mm}
\begin{example}\label{ex:5}
The database $D$ has relations $\nit{Client(ID,}$ $\nit{A,M)}$,
with key $\nit{ID}$, attributes $A$ for age and $M$ for  amount of
money; and $\nit{Buy(ID,I,P)}$, with key $\{\nit{ID},I\}$, $I$ for
items, and $P$ for prices. We have denials ~$\IC_1\!:~ \forall
\nit{ID}, P,A,M \neg($ $\nit{Buy(ID},I,P), \nit{Client(ID},A,M),A
< 18,$ $ P>25)$ ~and ~$\IC_2\!: \forall \nit{ID},A,M \neg($
$\nit{Client}($ $\nit{ID},A,M), A < 18,$ $M
>50)$, requiring that people younger than 18 can-

\vspace{-4mm}
\begin{multicols}{2}
\noindent $D$:\vspace*{-.4cm}
\begin{center}
{\small\begin{tabular}{c|c|c|c||c|} \hline
\multicolumn{1}{|c|}{~\bf Client~ ~}&~\bf ID~&~\bf A ~&~\bf
M~&\\
\hline & 1 & 15 & 52 & $t_1$\\
\cline{2-5} & 2 & 16 & 51 & $t_2$\\
\cline{2-5} & 3 & 60 & 900 & $t_3$\\\cline{2-5}
 \hline
\multicolumn{1}{|c|}{ \bf~Buy ~~~\,\,}&~ \bf ID\,&~\bf I ~&~\bf P
~&~
\\
\hline & 1 & CD & 27 &$t_4$ \\
\cline{2-5} & 1 & DVD & 26 &$t_5$ \\
\cline{2-5} & 3 & DVD & 40 & $t_6$\\
\cline{2-5}
\end{tabular}}
\end{center}

\vspace{-4mm}\noindent    not spend more than 25 on one item nor
spend more than 50 in the store. We added an extra column in the
tables with a label for  each tuple. $IC_1$ is violated by
\{$t_1$,$t_4$\} and \{$t_1$,$t_5$\}; and $\IC_2$ by \{$t_1$\} and
\{$t_2$\}. We have two LS-fixes (the modified version of tuple
$t_1$ is $t_1'$, etc.), with distances $\Delta(D,D')=$
\end{multicols}

\noindent $D'$: \hspace{5.3cm} $D''$:

\vspace*{-.38cm}  {\small
 ~~~\begin{tabular}{c|c|c|c||c|} \hline
\multicolumn{1}{|c|}{~\bf Client' ~}&~\bf ID~&~\bf A ~&~\bf
M~&\\
\hline & 1 & 15 & 50 & $t_1'$ \\
\cline{2-5} & 2 & 16 & 50 & $t_2$$'$\\
\cline{2-5} & 3 & 60 & 900 & $t_3$\\\cline{2-5} \hline
\multicolumn{1}{|c|}{ \bf~Buy' ~~\,\,}&~ \bf ID\,&~\bf I ~&~\bf P
~&~
\\
\hline & 1 & CD & 25 &$t_4$$'$ \\
\cline{2-5} & 1 & DVD & 25 &$t_5$$'$ \\
\cline{2-5} & 3 & DVD & 40 & $t_6$\\
\cline{2-5}
\end{tabular}

\vspace{-3.4cm}\hspace*{6.4cm}\begin{tabular}{c|c|c|c||c|} \hline
\multicolumn{1}{|c|}{~\bf Client'' ~}&~\bf ID~&~\bf A ~&~\bf
M~&\\
\hline & 1 & 18 & 52 &  $t_1$$''$ \\
\cline{2-5} & 2 & 16 & 50 & $t_2$$''$\\
\cline{2-5} & 3 & 60 & 900 & $t_3$\\\cline{2-5} \hline
\multicolumn{1}{|c|}{ \bf~Buy'' ~~\,\,}&~ \bf ID\,&~\bf I ~&~\bf P
~&~\\
\hline & 1 & CD & 27 &$t_4$ \\
\cline{2-5} & 1 & DVD & 26 &$t_5$ \\
\cline{2-5} & 3 & DVD & 40 & $t_6$\\
\cline{2-5}
\end{tabular}
}

\vspace*{.2cm} \noindent $ 2^2+1^2+2^2+1^2=10$, and
$\Delta(D,D'')= 3^2+1^2=10$. We can see that a global fix may not
be the result of applying ``local" minimal fixes to
tuples.\boxtheorem
\end{example}
\vspace{-2mm}The built-in atoms in  linear denials determine a
solution space for fixes as an intersection of semi-spaces, and
LS-fixes can be found at its ``borders" (c.f. previous example and
Proposition A.\ref{prop:prelimits} in Appendix \ref{sec:proof}).
It is easy to construct examples with an exponential number of
fixes. For the kind of fixes and ICs we are considering, it is
possible that no fix exists, in contrast to \cite{ABC99,ABC03},
where, if the set of ICs is consistent as a set of logical
sentences, a fix for a database always exist.

\vspace{-1mm}
\begin{example} \label{ex:nonexist}
$R(X,Y)$ has  key $X$ and  fixable $Y$. $\IC_1 = \{\forall X_1X_2Y
\neg (R(X_1,Y),$ $ R(X_2,$ $Y), X_1 \!=\!1, X_2\! =\! 2)$,
\!$\forall X_1X_2 Y \neg (R(X_1,$ $Y),$ $ R(X_2, Y),
X_1\!=\!1,X_2\! =\! 3)$, $\forall X_1X_2 Y \neg ($ $R(X_1,Y),
R(X_2, Y), \!X_1\!=\!2, \!X_2\!=\! 3)$, \!$\forall XY \neg
(R(X,Y), Y\!\!>\!3)$, \!$\forall XY \neg ($ $R(X,Y),Y\!<\!2)\}$ is
consistent. The first three ICs force  $Y$ to be different in
every tuple. The last two ICs require $2 \leq Y \leq 3$. The
inconsistent database $R = \{(1,-1), (2,1),$ $(3,5)\}$ has no fix.
Now, for $\IC_2$ with
 $\forall X,Y \neg( R(X,Y),$ $ Y > 1)$ and
$\nit{sum}(Y) = 10$, any database with less than 10 tuples has no
fixes. \boxtheorem
\end{example}

\vspace{-4mm}
\begin{proposition}\label{lemma:existence} \em
If  $D$ has a fix wrt $\IC$, then it also has an LS-fix wrt $\IC$.
\boxtheorem
\end{proposition}

\vspace{-6mm}
\section{Decidability and Complexity}\label{sec:Complexity}

\vspace{-1mm}In applications where fixes are based on changes of
numerical values, computing concrete fixes is a relevant problem.
In databases containing census forms, correcting the latter before
doing statistical processing is a common problem \cite{franconi}.
In databases with experimental samples, we can fix certain
erroneous quantities as specified by linear ICs. In these cases,
the fixes are relevant objects to compute explicitly, which
contrasts with CQA \cite{ABC99}, where the main motivation for
introducing repairs is to formally characterize the notion of a
consistent answer to a query as an answer that remains under all
possible repairs. In consequence, we now consider some decision
problems related to existence and verification of LS-fixes, and to
CQA under different semantics.

\begin{definition}\label{def:problems} \em
For an instance $D$ and a set $\IC$ of ICs:\\(a)
$\nit{Fix}(D,\IC):=$ $\{D^\prime ~|~D^\prime~\mbox{ is an }$
$\mbox{LS-fix of } {\it D}$ $\mbox{wrt } \IC\}$, the {\em fix
checking problem}.\\(b) $\nit{Fix}(\IC) := \{(D,D')$ $|~ D' \in
\nit{Fix}(D,\IC)\}$.\\(c)~$\nit{NE}(\IC) := \{D~|$
$\nit{Fix}(D,\IC)$ $ \neq \emptyset\}$, for {\em non-empty} set of
fixes, i.e. the problem of {\em checking existence of
LS-fixes}.\\(d)~$\nit{NE} := \{ (D,\IC) ~|~ \nit{Fix}(D,\IC) \neq
\emptyset\}$.\\(e) $\nit{DFP}(\IC) \!:=\!\! \{(D,k)| \mbox{ there
is } D^\prime \in Fix(D,\IC) \mbox{ with } \Delta(D,D') \leq k\}$,
the {\em database fix problem}, i.e. the problem of checking
existence of LS-fixes within a given positive distance $k$.\\
(f) $\nit{DFOP}(\IC)$ is the optimization problem of finding the
minimum distance from an LS-fix wrt $\IC$ to a given input
instance.\boxtheorem
\end{definition}

\begin{definition}  \label{def:semantics} \em Let $D$ be a database, $\IC$ a set ICs,
and $Q$ a conjunctive query\footnote{Whenever we say just
``conjunctive query" we understand it
is a non aggregate query.}.\\
(a) A ground tuple $\bar{t}$ is a {\em consistent answer} to
$Q(\bar{x})$ under the:~ (a1)
 {\em skeptical semantics} if for every $D' \in \nit{Fix}(D,\IC)$, $D' \models
    Q(\bar{t})$.~
(a2)  {\em brave semantics} if there exists $D' \in
\nit{Fix}(D,\IC)$ with $D' \models
    Q(\bar{t})$.~
  (a3) {\em majority semantics} if
 $|\{ D^\prime ~|~ D^\prime \in \nit{Fix}(D,\IC)$
and $D^\prime \models$ $Q(\bar{t})\}|$ $>$
 $|\{D^\prime ~|~ D^\prime \in$ $\nit{Fix}(D,\IC)$
and $D^\prime \not \models Q(\bar{t})\}|$.\\ (b) That $\bar{t}$ is
a consistent answer to $Q$ in $D$ under semantics $\cal S$ is
denoted by $D \models_{\cal S} Q[\bar{t}]$. If $Q$ is ground and
$D \models_{\cal S} Q$, we say that {\em yes}~ is a consistent
answer, meaning that $Q$ is true in the fixes of $D$ according to
semantics $S$. $\nit{CA}(Q,D,\IC,{\cal S})$ is the set of
consistent answers to $Q$ in $D$ wrt \IC\ under semantics ${\cal
S}$. For ground $Q$, if $\nit{CA}(Q,D,\IC,{\cal S}) \neq
\{\nit{yes}\}$, $\nit{CA}(Q,D,\IC,{\cal S}) := \{\nit{no}\}$.\\
(c) $\nit{CQA}(Q,\IC,{\cal S}) := \{(D,\bar{t}) ~|~ \bar{t} \in
\nit{CA}(Q,D,\IC,{\cal S})\}$ is the decision {\em problem of
consistent query answering}, of  checking consistent answers.
\boxtheorem
\end{definition}
\begin{proposition}\label{lemma:reduc} \em
$\nit{NE}({\IC})$ can be reduced in polynomial time to the
complements of $\nit{CQA}(\nit{False},\IC,$ $\nit{Skeptical})$ and
$\nit{CQA}(\nit{True},\IC, \nit{Majority})$, where $\nit{False},
\nit{True}$ are ground queries that are always false, resp. true.
\boxtheorem
\end{proposition}
In Proposition \ref{lemma:reduc}, it suffices for queries
$\nit{False}, \nit{True}$ to be false, resp. true, in all
instances that share the key values with the input database. Then,
 they can be represented by $\exists Y R(\bar{c},Y)$, where $\bar{c}$ are
not (for {\em False}), or are (for {\em True})  key values  in the
original instance.

\begin{theorem} \label{theo:exist} \em  Under extended linear
denials and complex, filtering, multi-attri-bute, single-relation,
aggregation constraints, the problems $\nit{NE}$ of existence of
LS-fixes, and $\nit{CQA}$ under  skeptical or majority semantics
are undecidable. \boxtheorem
\end{theorem}
The result about $\nit{NE}$ can be proved by reduction from the
undecidable Hilbert's problem on solvability of diophantine
equations. For CQA,  apply Proposition \ref{lemma:reduc}. Here we
have the original database and the set of ICs as input parameters.
In the following we will be interested in data complexity, when
only the input database varies and the set of ICs is fixed
\cite{AHV95}.\vspace{-1mm}

\begin{theorem} \label{lemma:squa-member} \em For a fixed set
$\IC$~ of  linear denials: (a) Deciding if for an instance $D$
there is an instance $D'$ (with the same key values as $D$) that
satisfies \IC\ with $\Delta(D,D') \leq k$, with positive integer
$k$ that is part of the input, is in $\nit{NP}$. (b)
$\nit{DFP}(\IC)$ is $\nit{NP}$-complete. (c.f. Definition
\ref{def:problems}(e)) \boxtheorem
\end{theorem}
\vspace{-1mm}By Proposition \ref{lemma:existence}, there is a fix
for $D$ wrt $\IC$ at a distance $\leq k$ iff there is an LS-fix at
a distance  $\leq k$. Part (b) of Theorem \ref{lemma:squa-member}
follows from part (a) and a reduction of {\em Vertex Cover} to
$\nit{DFP}(\IC_0)$, for a fixed set of denials $\IC_0$. By Theorem
\ref{lemma:squa-member}(a), if there is a fix at a distance $\leq
k$, the minimum distance to $D$ for a fix can be found by binary
search in $log(k)$ steps. Actually, if an LS-fix exists, its
square distance to $D$ is polynomially bounded by the size of $D$
(c.f. proof of Theorem \ref{cor:finNPcomp}). Since $D$ and a fix
have the same number of tuples, only the size of their values in a
fix matter, and they are constrained by a fixed set of linear
denials and the condition of minimality.
\begin{theorem} \label{cor:finNPcomp} \em For a fixed set $\IC$ of extended linear
denials: (a) The problem $\nit{NE}(\IC)$ of deciding if an
instance has  an LS-fix wrt $\IC$~ is $\nit{NP}$-complete, and (b)
$\nit{CQA}$ under the skeptical and the majority semantics is
$\nit{coNP}$-hard. \boxtheorem
\end{theorem}
For hardness in (a), (b) in Theorem \ref{cor:finNPcomp}, linear
denials are good enough. Membership in (a) can be obtained for any
fixed finite set of extended denials. Part (b) follows from part
(a). The latter uses a reduction from {\em 3-Colorability}.
\begin{theorem}\label{theo:checking} \em
For a fixed set $\IC$ of extended linear denials: (a) The problem
$\nit{Fix}(\IC)$ of checking if an instance is an LS-fix is
$\nit{coNP}$-complete, and (b) $\nit{CQA}$ under skeptical
semantics is in $\Pi^P_2$, and, for ground atomic queries,
$\Delta_2^P$-hard. \boxtheorem
\end{theorem}
Part (a) uses {\em 3SAT}. Hardness in (b) is obtained by reduction
from a $\Delta^P_2$-complete decision version of the problem of
searching for the lexicographically {\em Maximum 3-Satisfying
Assignment} ({\em M3SA}): Decide if the last variable takes value
$1$ in it \cite[Theo. 3.4]{krentel}. Linear denials suffice. Now,
by  reduction from the {\em Vertex Cover Problem}, we obtain
\begin{theorem} \label{prop:minCover} \em For aggregate comparison queries using
$\nit{sum}$, $\nit{CQA}$ under linear denials and brave semantics
is $\nit{coNP}$-hard. \boxtheorem
\end{theorem}

\section{Approximation for the Database Fix Problem} \label{sec:approx}
We consider the problem of finding a good approximation for the
general optimization problem $\nit{DFOP}(\IC)$.
\begin{proposition}\label{lem:dfp-snphard} \em For a fixed set of
linear denials $\IC$, $\nit{DFOP}(\IC)$ is $\nit{MAXSNP}$-hard.
\boxtheorem
\end{proposition}
 This result is obtained by establishing an
$L$-reduction to $\nit{DFOP}(\IC)$
 from the $\nit{MAXSNP}$-complete \cite{papYannak91,papadimitriou94}
\emph{B-Minimum Vertex Cover Problem}, i.e. the  vertex cover
minimization problem for graphs of bounded degree \cite[Chapter
10]{hochbaum}. As an immediate consequence, we obtain that
$\nit{DFOP}(\IC)$ cannot be uniformly approximated within
arbitrarily small constant factors \cite{papadimitriou94}.
\begin{corollary} \em Unless $P \!= \!\nit{NP}$, there is no {\em Polynomial
Time Approximation Schema}  for $\nit{DFOP}$. \boxtheorem
\end{corollary}
This negative result does not preclude the possibility of finding
an efficient algorithm for approximation within a constant factor
for $\nit{DFOP}$. Actually, in the following we do this for a
restricted but still useful class of denial constraints.

\subsection{Local denials}

\begin{definition} \label{def:local} \em A set of linear denials $\IC$~ is
\emph{local} if: (a) Attributes participating in equality atoms
between attributes or in joins are all rigid; (b) There is a
built-in atom with a fixable attribute in each element of $\IC$;
(c) No attribute $A$ appears in $\IC$ both in comparisons of the
form $A < c_1$ and $A
> c_2$.\footnote{To check condition (c), $x\leq c,~ x\geq c,~ x\neq c$ have
to be expressed using $<,>$, e.g. $x\leq c$ by $x<
c+1$.}\boxtheorem
\end{definition}
In Example \ref{ex:5}, $\IC$ is local. In Example
\ref{ex:nonexist}, $\IC_1$ is not local. Local constraints  have
the property that by doing local fixes, no new inconsistencies are
generated, and there is always an LS-fix wrt to them (c.f.
Proposition A.\ref{prop:itworks} in Appendix \ref{sec:proof}).
Locality is a sufficient, but not necessary condition for
existence of LS-fixes as can be seen from the database
$\{P(a,2)\}$, with the first attribute as a key and non-local
denials $\neg(P(x,y), y <3), \neg(P(x,y), y
>5)$, that  has the LS-fix $\{P(a,3)\}$.
\begin{proposition}\label{prop:still} \em
For the class of local denials, $\nit{DFP}$  is
$\nit{NP}$-complete, and $\nit{DFOP}$ is $\nit{MAXSNP}$-hard.
\boxtheorem
\end{proposition}
This proposition tells us that the problem of finding good
approximations in the case of local denials is still relevant.

\begin{definition}\label{def:inctupset} \em
A set $I$ of database tuples from  $D$ is a {\em violation set}
for $\nit{ic} \in \IC$ if $I\not \models \nit{ic}$, and for every
$I' \subsetneqq I$, $I' \models \nit{ic}$. ${\cal
I}(D,\nit{ic},t)$ denotes the set of violation sets for $\nit{ic}$
that contain tuple $t$. \boxtheorem
\end{definition}
A violation set $I$ for $\nit{ic}$ is a minimal set of tuples that
simultaneously participate in the violation of $\nit{ic}$.

\begin{definition}\label{def:locfix} \em
Given an instance $D$ and ICs $\IC$, a {\em local fix} for  $t \in
D$, is a tuple $t'$ with: (a) the same values for the rigid
attributes as $t$; (b) $S(t,t') := \{I~|~ \mbox{there is }\nit{ic}
\in \IC,~ I \in {\cal I}(D,\nit{ic},t)\mbox{ and }$ $((I
\smallsetminus \{t\}) \cup \{t'\}) \models \nit{ic}\} ~\neq~
\emptyset$; ~and (c) there is no tuple $t''$ that simultaneously
satisfies (a), $S(t,t'') = S(t,t')$, and $\Delta(\{t\},\{t''\})
\leq \Delta(\{t\},\{t'\})$, where $\Delta$ denotes quadratic
distance. \boxtheorem
\end{definition}
$S(t,t')$ contains the violation sets that include $t$ and are
solved by replacing $t'$ for $t$. A local fix $t'$ solves some of
the violations due to $t$ and minimizes the distance to $t$.

\subsection{Database fix problem as a set cover problem}

For a fixed set $\IC$ of local denials, we can solve an instance
of $\nit{DFOP}$ by transforming it into an instance of the
 {\em Minimum Weighted Set Cover Optimization
Problem} ($\!\nit{MWSCP}\!$). This problem is $\nit{MAXSNP}$-hard
\cite{lund,papadimitriou94}, and its general approximation
algorithms are within a logarithmic factor \cite{lund,chvatal}. By
concentrating on local denials, we will be able to generate a
version of the $\nit{MWSCP}$ that can be approximated within a
constant factor (c.f. Section \ref{sec:appro}).
\begin{definition} \em
For a database $D$ and a set $\IC$ of local denials, ${\cal
G}(D,\IC) = (T,H)$ denotes the {\em conflict hypergraph} for $D$
wrt $\IC$ \cite{chomickiIC}, which has in the set $T$ of vertices
the database tuples, and in the set $H$ of hyperedges, the
violation sets for elements $\nit{ic} \in \IC$. \boxtheorem
\end{definition}
Hyperedges in $H$ can be labelled with the corresponding
$\nit{ic}$, so that different hyperedges may contain the same
tuples. Now we build an instance of $\!\nit{MWSCP}\!$.
\begin{definition} \em For
a database $D$ and a set $\IC$ of local denials, the instance
$(U,{\cal S}, w)$ for the $\!\nit{MWSCP}\!$, where $U$ is the
underlying set, ${\cal S}$ is the set collection, and $w$ is the
weight function, is given by: (a)  $U := H$, the set of hyperedges
of ${\cal G}(D,\IC)$. (b) ${\cal S}$ contains the $S(t,t')$, where
$t'$ is a local fix for a tuple $t \in D$. (c) $w(S(t,t')) :=
\Delta(\{t\},\{t'\})$. \boxtheorem
\end{definition}
It can be proved that the $S(t,t')$ in this construction are non
empty, and that ${\cal S}$ covers $U$ (c.f. Proposition
A.\ref{prop:itworks} in Appendix \ref{sec:proof}).

If for the instance $(U,{\cal S},w)$ of $\nit{MWSCP}$ we find  a
minimum weight cover $\cal C$, we could think of constructing a
fix by replacing each inconsistent tuple $t \in D$ by a local fix
$t'$ with $S(t,t') \in {\cal C}$. The problem is that there might
be more than one $t'$ and the key dependencies would not be
respected. Fortunately, this problem can be circumvented.
\begin{definition} \label{def:star} \em
Let ${\cal C}$ be a cover for instance $(U,{\cal S},w)$ of the
$\nit{MWSCP}$ associated to $D, \IC$.~ (a) ${\cal C}^\star$ is
obtained from ${\cal C}$ as follows: For each tuple $t$ with local
fixes $t_1, \dots, t_n$, $n
>1$, such that $S(t,t_i) \in {\cal C}$, replace in
$\cal C$ all the $S(t,t_i)$ by a single $S(t,t^\star)$, where
$t^\star$ is such that $S(t,t^\star)= \bigcup_{i=1}^n S(t,t_i)$.~
(b) $D({\cal C})$ is the database instance obtained from $D$ by
replacing $t$ by $t'$ if $S(t,t') \in {\cal C}^\star$. \boxtheorem
\end{definition}
It holds (c.f. Proposition A.\ref{prop:LFcombined} in Appendix
\ref{sec:proof}) that  such an $S(t,t^\star) \in {\cal S}$  exists
in part (a) of Definition \ref{def:star}. Notice that there, tuple
$t$ could have other $S(t,t')$ outside ${\cal C}$. Now we can show
that the reduction to $\nit{MWSCP}$ keeps the value of the
objective function.
\begin{proposition} \label{prop:keep} \em
If ${\cal C}$ is an optimal cover for instance $(U,{\cal S},w)$ of
the $\nit{MWSCP}$ associated to $D, \IC$, then $D({\cal C})$ is an
LS-fix of $D$ wrt $\IC$, and $\Delta(D,D({\cal C})) = w({\cal C})
= w({\cal C}^*)$. \boxtheorem
\end{proposition}
\begin{proposition} \label{prop:allLSfix} \em For every LS-fix $D'$ of
 $D$ wrt a set of local denials $\IC$, there exists an optimal
cover $\cal C$ for the associated instance $(U,{\cal S},w)$ of the
$MWSCP$, such that $D'=D({\cal C})$. \boxtheorem
\end{proposition}
\begin{proposition} \label{theo:ApproxNonCon} \em
The transformation of $\nit{DFOP}$ into $\nit{MWSCP}$, and  the
construction of database instance $D({\cal C})$ from a cover
${\cal C}$ for $(U,{\cal S},w)$ can be done in polynomial time in
the size of $D$. \boxtheorem
\end{proposition}
We have established that the transformation of $\nit{DFOP}$ into
$\nit{MWSCP}$ is an $L$-reduction \cite{papadimitriou94}.
Proposition \ref{theo:ApproxNonCon} proves, in particular, that
the number of violation sets $S(t,t')$ is polynomially bounded by
the size of the original database $D$.

\begin{example} (example \ref{ex:5} continued) \label{ex:setco} We illustrate the
reduction from $\nit{DFOP}$ to $\nit{MWSCP}$. The violation sets
are \{$t_1$,$t_4$\} and \{$t_1$,$t_5$\} for $\nit{IC}_1$ and
\{$t_1$\} and \{$t_2$\} for $\nit{IC}_2$. The figure shows the
hypergraph. For the $\nit{MWSCP}$ instance, we need the local
fixes. Tuple $t_1$ has two local fixes $t_1'=(1,15,50)$, that
solves the violation set $\{t_1\}$ of $\IC_2$ (hyperedge B), and
$t_1''=(1,18,52)$, that solves the violation sets $\{t_1,t_4\}$
and $\{t_1,t_5\}$ of $\nit{IC}_1$, and $\{t_1\}$ of $\nit{IC}_2$
(hyperedges A,B, C), with weights  $4$ and $9$, resp. $t_2$, $t_4$
and $t_5$ have one local fix each corresponding to: $(2,16,50)$,
\noindent $(1,\nit{CD},25)$ and $(1,\nit{DVD},25)$, resp. ~The
consistent tuple $t_3$ has no local fix.

\begin{multicols}{2}

\includegraphics[width=5cm]{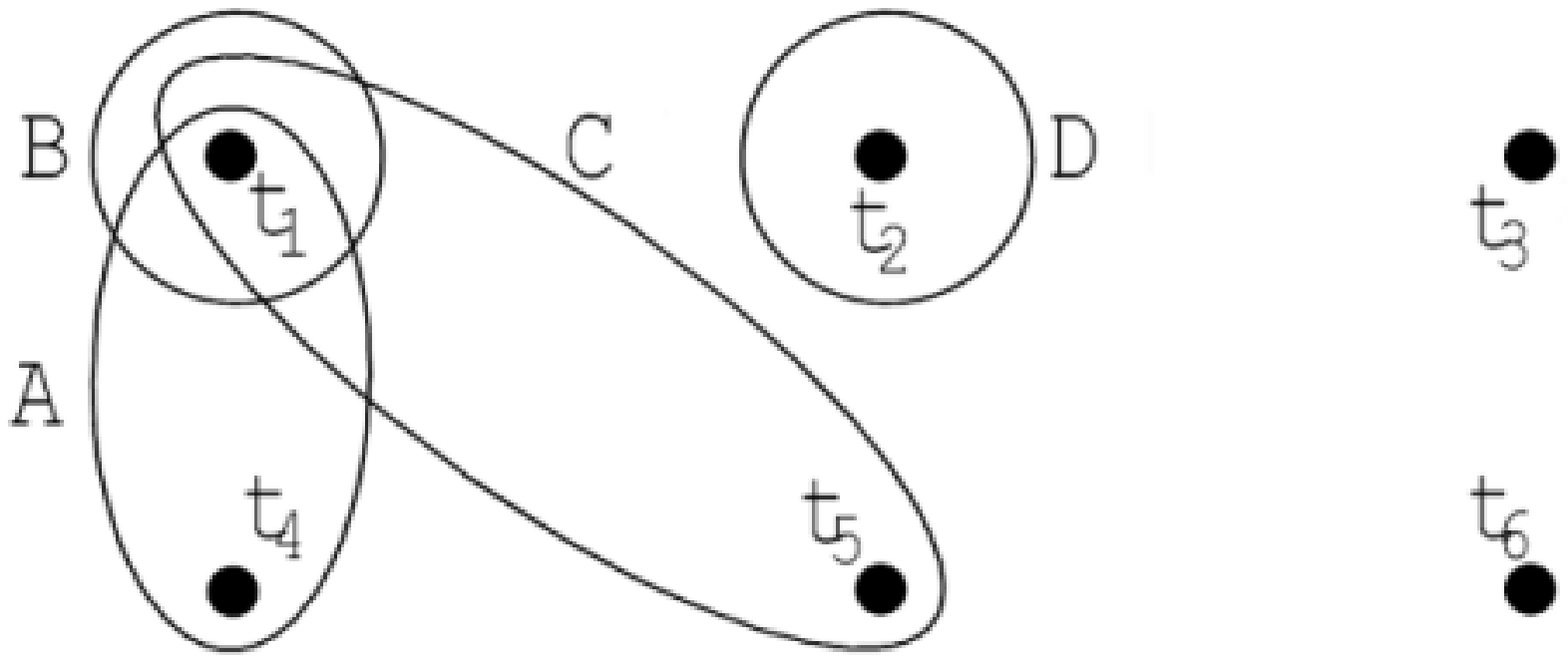}\\

\begin{tabular}{|l|c|c|c|c|c|}
\hline Set & $S_1$&$S_2$ &$S_3$ &$S_4$ & $S_5$\\\hline
Local Fix  & $t_1$' & $t_1$'' & $t_2$' &  $t_4$' & $t_5$' \\
Weight  & 4  & 9   & 1  &  4  & 1  \\
\hline
Hyperedge A & 0  & 1   & 0  &  1  & 0  \\
Hyperedge B & 1  & 1   & 0  &  0  & 0  \\
Hyperedge C & 0  & 1   & 0  &  0  & 1  \\
Hyperedge D & 0  & 0   & 1  &  0  & 0 \\
\hline
\end{tabular}
\end{multicols}

\vspace{1mm}\noindent The $\nit{MWSCP}$ instance is shown in the
table, where the elements are  rows and the sets (e.g. $S_1=
S(t_1,t_1')$), columns. An entry $1$ means that the set contains
the corresponding element; and a $0$, otherwise. There are two
minimal covers, both with weight $10$: ${\cal C}_1 = \{S_2, S_3\}$
and ${\cal C}_2 = \{S_1, S_3, S_4, S_5\}$. $D({\cal C}_1)$ and
$D({\cal C}_2)$ are the two fixes for this problem. \boxtheorem
\end{example}
If we apply the transformation to Example \ref{ex:nonexist}, that
had non-local set of ICs and no repairs, we will find that
instance $D(\cal C)$, for ${\cal C}$ a set cover, can be
constructed as above, but it does not satisfy the flexible ICs,
because changing inconsistent tuples by their local fixes solves
only the initial inconsistencies, but new inconsistencies are
introduced.

\subsection{Approximation via set cover optimization}
\label{sec:appro}  Now that we have transformed the database fix
problem into a weighted set cover problem, we can apply
approximation algorithms for the latter. We know, for example,
that using a greedy algorithm, $\nit{MWSCP}$ can be approximated
within a factor $\nit{log}(N)$, where $N$ is the size of the
underlying set $U$ \cite{chvatal}. The approximation algorithm
returns not only an approximation $\hat{w}$ to the optimal weight
$w^o$, but also a -non necessarily optimal- cover $\hat{\cal C}$
for problem $(U,{\cal S},w)$. As in Definition \ref{def:star},
$\hat{\cal C}$ can be used to generate via $(\hat{\cal C})^\star$,
a fix $D(\hat{\cal C})$ for $D$ that may not be LS-minimal.

\begin{example} (examples \ref{ex:5} and \ref{ex:setco} continued) We show how to
to compute a solution to this particular instance of $\nit{DFOP}$
using the greedy approximation algorithm for $\nit{MWSCP}$
presented in \cite{chvatal}. We start with ${\cal \hat{C}}:=
\emptyset$, $S_i^0 := S_i$; and  we add to $\cal C$ the $S_i$ such
that $S_i^0$ has the maximum {\em contribution ratio}
$|S_i^0|/w(S_i^0)$. The alternatives are $|S_1|/w(S_1)=1/4$,
$|S_2|/w(S_2)=3/9$, $|S_3|/w(S_3)=1$, $|S_4|/w(S_4)=1/4$ and
$|S_5|/w(S_5)=1$. The ratio is maximum for $S_3$ and $S_5$, so we
can add any of them to $\hat{\cal C}$. If we choose the first, we
get ${\cal \hat{C}}= \{S_3\}$. Now we compute the $S_i^1:= S_i^0
\smallsetminus S_3^0$, and choose again an $S_i$ for $\hat{\cal
C}$ such that $S_i^1$ maximizes the contribution ratio. Now $S_5$
is added to $\hat{\cal C}$, because $S_5^1$ gives the maximum. By
repeating this process until we get all the elements of $U$
covered, i.e. all the $S_i^k$ become empty at some iteration point
$k$, we finally obtain $\hat{\cal C}= \{S_3, S_5, S_1,S_4\}$. In
this case ${\cal \hat{C}}$ is an optimal cover and therefore,
$D({\cal \hat{C}})$ is exactly an LS-fix, namely $D'$ in Example
\ref{ex:5}. Since this is an approximation algorithm, in other
examples the cover obtained might not be optimal. \boxtheorem
\end{example}
\begin{proposition} \label{prop:ApproxCover}\em Given database instance $D$ with local ICs $\IC$,  the
database instance $D(\hat{\cal C})$ obtained from the approximate
cover $\hat{\cal C}$ is a fix and it holds $\Delta(D,D(\hat{\cal
C})) \leq log(N) \times \Delta(D,D')$, where $D'$ is any LS-fix of
D wrt $\IC$ and $N$ is the number of of violation sets for $D$ wrt
$\IC$. \boxtheorem
\end{proposition}
In consequence, for any set $\IC$ of local denials, we have a
polynomial time approximation algorithm that solves
$\nit{DFOP}(\IC)$  within an $O(\nit{log}(N))$ factor, where $N$
is the number of violation sets for $D$ wrt $\IC$. As mentioned
before, this number $N$, the number of hyperedges in ${\cal G}$,
is polynomially bounded by $|D|$ (c.f. Proposition
\ref{theo:ApproxNonCon}). $N$ may be small if the number of
inconsistencies or the number of database atoms in the ICs are
small, which is likely the case in real applications.

However, in our case we can get even better approximations via a
cover $\hat{\cal C}$ obtained with an approximation algorithms for
the special case of the $\nit{MWSCP}$ where the number of
occurrences of an element of $U$ in elements of ${\cal S}$ is
bounded by a constant. For this case of the $\nit{MWSCP}$ there
are approximations within a constant factor based on ``linear
relaxation" \cite[Chapter 3]{hochbaum}. This is clearly the case
in our application, being $m \times |{\cal F}| \times |\nit{IC}|$
a constant bound (independent from $|D|$) on the frequency of the
elements, where $m$ is the maximum number of database atoms in an
IC.
\begin{theorem} \em
There is an approximation algorithm that, for a given database
instance $D$ with local ICs $\IC$, returns a fix $D(\hat{\cal C})$
such that $\Delta(D,D(\hat{\cal C})) \leq c \times \Delta(D,D')$,
where $c$ is a constant and $D'$ is any LS-fix of $D$. \boxtheorem
\end{theorem}

\section{One Atoms
Denials and Conjunctive Queries}\label{sec:tractable}

 In this section we concentrate on the common case of
{\em one database atom denials} (1AD), i.e. of the form $\forall
\neg(A, B)$, where atom $A$ has a predicate in ${\cal R}$, and $B$
is a conjunction of built-in atoms. They capture range
constraints; and census data is usually stored in single relation
schemas \cite{franconi}.

For 1ADs, we can identify tractable cases for $\nit{CQA}$ under
LS-fixes by reduction to $\nit{CQA}$ for (tuple and set-theoretic)
repairs of the form introduced in \cite{ABC99} for key
constraints. This is because each violation set (c.f. Definition
\ref{def:inctupset}) contains one tuple, maybe with several local
fixes, but all sharing the same key values; and then the problem
consists in choosing one from different tuples with the same key
values (c.f. proof  of Theorem \ref{theo:forest}). The
transformation preserves consistent answers to both ground and
open queries.

The ``classical" -tuple and set oriented- repair problem as
introduced in \cite{ABC99} has been studied in detail for
functional dependencies in \cite{chomickiIC,fuxman}. In
particular, for tractability of $\nit{CQA}$ in our setting, we can
use results  and algorithms obtained in \cite{fuxman} for the
classical framework.

The {\em join graph ${\cal G}(Q)$} \cite{fuxman} of a conjunctive
query $Q$ is a directed graph, whose vertices are the database
atoms in $Q$. There is an arc from $L$ to $L'$ if $L \neq L'$ and
there is a variable $w$ that occurs at the position of a non-key
attribute in $L$ and also occurs in $L'$. Furthermore, there is a
self-loop at $L$ if there is a variable that occurs at the
position of a non-key attribute in $L$, and at least twice in $L$.

When $Q$ does not have repeated relations symbols, we write $Q \in
{\cal C}_{\!\it Tree}$ if ${\cal G}(Q)$ is a forest and every
non-key to key join of $Q$ is full i.e. involves the whole key.
Classical $\nit{CQA}$ is tractable for queries in ${\cal C}_{\!\it
Tree}$  \cite{fuxman}.
\begin{theorem}\label{theo:forest} \em For a fixed
set of 1ADs and queries in $C_{\!\nit{Tree}}$, consistent query
answering under LS-fixes is in $\nit{PTIME}$. \boxtheorem
\end{theorem}
We may define that a aggregate conjunctive query belongs to
$C_{\!\nit{Tree}}$ if its underlying non-aggregate conjunctive
query, i.e. its NAM (c.f. Section \ref{sec:prel}) belongs to
$C_{\!\nit{Tree}}$. Even for 1ADs, with simple comparison
aggregate queries with $\nit{sum}$, tractability is lost under the
brave  semantics.
\begin{proposition}\label{prop:possaggregsimple} \em
For a fixed set of 1ADs,  and for aggregate queries that are in
$C_{\!\nit{Tree}}$ or acyclic, $\nit{CQA}$ is $\nit{NP}$-hard
under the brave semantics. \boxtheorem
\end{proposition}
For  queries $Q$ returning numerical values, which is common in
our framework, it is natural to use the {\em range semantics} for
$\nit{CQA}$,  introduced in \cite{ABC03} for scalar aggregate
queries and functional dependencies under classical repairs. Under
this semantics, a consistent answer is the pair consisting of the
{\em min-max} and {\em max-min} answers, i.e. the supremum and the
infimum, resp., of the set of answers to $Q$ obtained from
LS-fixes. The $\nit{CQA}$ decision problems under range semantics
consist in determining if a numerical query $Q$, e.g. an aggregate
query, has its answer $\leq k_1$ in every fix ({\em min-max}
case), or $\geq k_2$ in every fix ({\em max-min} case).
\begin{theorem}\label{theo:range} \em
For each of the aggregate functions {\em sum, count distinct}, and
{\em average}, there is a fixed set of 1ADs and a fixed aggregate
acyclic conjunctive query, such that $\nit{CQA}$ under the range
semantics is ${\it NP}$-hard. \boxtheorem
\end{theorem}
For the three aggregate functions one 1AD suffices. The results
for {\em count distinct} and {\em average} are obtained by
reduction from $\nit{MAXSAT}$ \cite{papadimitriou94} and
$\nit{3SAT}$, resp. For {\em sum}, we use a reduction from the
{\em Independent Set Problem} with bounded degree 3
\cite{GareyJohnsonStockmayer}. The general {\em Independent Set
Problem} has bad approximation properties \cite[Chapter
10]{hochbaum}. The {\em Bounded Degree Independent Set} has
efficient approximations within a constant factor that depends on
the degree \cite{hr-94}.
\begin{theorem}\label{theo:approx-sum} \em For any
set of 1ADs and conjunctive query with {\em sum} over a
non-negative attribute, there is a polynomial time approximation
algorithm with a constant factor for {\em CQA} under {\em min-max}
range semantics. \boxtheorem
\end{theorem}
\vspace{-1mm}The factor in this theorem depends upon the ICs and
the query, but not on the size of the database. The acyclicity of
the query is not required. The algorithm is based on a reduction
of our problem to satisfying a subsystem with maximum weight of a
system of weighted algebraic equations over the Galois field with
two elements $\nit{GF}[2]$ (a generalization of problems in
\cite{garey,vazirani}), for which a polynomial time  approximation
similar to the one for {\em MAXSAT} can be given \cite{vazirani}.

\vspace{-1mm}\section{Conclusions}\label{sec:conclusions}
\vspace{-1mm} We have shown that fixing numerical values in
databases  poses many new computational challenges that had not
been addressed before in the context of consistent query
answering. These problems are particularly relevant in census like
applications, where the problem of {\em data editing} is a common
and difficult task (c.f. {\tt
http://www.unece.org/stats/documents/2005.05.sde.htm}). Also our
concentration on aggregate queries is particularly relevant for
this kind of statistical applications. In this paper we have just
started to investigate some of the many problems that appear in
this context, and several extensions are in development. We
concentrated on integer numerical values, which provide a useful
and challenging domain. Considering real numbers in fixable
attributes opens many new issues, requires different approaches;
and is a subject of ongoing research.

\vspace{-.5mm}The framework established in this paper could be
applied to qualitative attributes with an implicit linear order
given by the application.  The result we have presented for
fixable attributes that are all equally relevant ($\alpha_{\!_A} =
1$ in Definitions \ref{def:dist} and \ref{def:LSrep}) should carry
over without much difficulty to the general case of arbitrary
weighted fixes. We have developed (but not reported here)
extensions to our approach that consider {\em minimum distribution
variation} LS-fixes that keep the overall statistical properties
of the database. We have also developed optimizations of the
approximation algorithm presented in Section \ref{sec:approx}; and
its implementation and experiments are ongoing efforts. More
research on the impact of aggregation constraints on LS-fixes is
needed.

\vspace{-.5mm}Of course, if instead of the $L_2$ distance, the
$L_1$ distance is used, we may get for the same database a
different set of (now $L_1$) fixes. The actual approximations
obtained in this paper change too. However, the general complexity
and approximability results should remain. They basically depend
on the fact that distance functions are non-negative, additive wrt
attributes and tuples,  computable in polynomial time, and
monotonically increasing. Another possible semantics could
consider an epsilon of error in the distance in such a way that
if, for example, the distance of a fix is $5$ and the distance to
another fix is $5.001$, we could take both of them as (minimal)
LS-fixes.

\vspace{-.5mm}Other open problems refer to cases of polynomial
complexity for linear denials with more that one database atom;
approximation algorithms for the {\em DFOP} for non-local cases;
and approximations to CQA for other aggregate queries.

\vspace{-.5mm}For related work, we refer to the literature on
consistent query answering (c.f. \cite{bookChapter} for a survey
and  references). Papers \cite{Wij03} and \cite{franconi} are the
closest to our work, because  changes in attribute values are
basic repair actions, but the peculiarities of numerical values
and quantitative distances between databases are not investigated.
Under the set-theoretic, tuple-based semantics,
\cite{chomickiIC,cali,fuxman} report on complexity issues for
conjunctive queries, functional dependencies and foreign key
constraints. A majority semantics was studied in \cite{mendelzon}
for database merging. Quite recent papers, but under semantics
different than ours, report research on fixing numerical values
under aggregation constraints \cite{flesca}; and heuristic
construction of repairs based on attribute values changes
\cite{bell}.

{\noindent  {\bf Acknowledgments:}~~ Research supported by NSERC,
CITO/IBM-CAS Student Internship Program, and EU projects: Sewasie,
Knowledge Web, and Interop. ~L. Bertossi is Faculty Fellow of  IBM
Center for Advanced Studies (Toronto Lab.).

}

\appendix
\section{Appendix}

\subsection{Proofs} \label{sec:proof}

Those auxiliary technical results that are stated in this
appendix, but not in the main body of the paper, are numbered in
the form ${\bf A.n}$, e.g. Lemma A.1.

\vspace{.3cm}\defproof{Proposition \ref{lemma:existence}}{Let
$\rho$ be the square distance between $D$ and $D^\prime$ in
Definition \ref{def:dist}. The circle of radius $\rho$ around $D$
intersects the non empty ``consistent" region that contains the
database instances with the same schema and key values as $D$ and
satisfy $\IC$. Since the circle has a finite number of instances,
the distance takes a minimum in the consistent region.}

 The following lemma proves
that if a tuple is involved in an inconsistency, the set of
constraints is consistent and there is at least one fixable
attribute in each integrity constraint, then there always exists a
local fix (see Definition \ref{def:locfix}) for it.

\begin{lemmaA} \label{lemma:existLF} For a database $D$ and a consistent set of linear denial
constraints $\IC$, where each constraint contains at least one
built-in involving a flexible constraint and there are equalities
or joins only between rigid attributes.  Then, for every tuple $t$
with at least one fixable attribute and at least one $\nit{ic}$ in
$\IC$, ${\cal I}(D,ic,t) \neq \emptyset$, there exists at least
one local fix $t'$ (see Definition \ref{def:locfix})\boxtheorem\\
\end{lemmaA}

\dproof{}{ Each constraint $\nit{ic} \in \IC $ has the form
$\forall \bar{x} \neg(P_1(\bar{x}),\ldots,$ $P_n(\bar{x}),$ $A_i <
c_i, A_j \geq c_j,A_k=c_k,A_l\neq c_l, \ldots)$ and can be
rewritten as a clause only with $<$, $>$ and $=$:
\vspace{-.25cm}\begin{equation} \forall \bar{x} (\neg
P_1(\bar{x})\vee \ldots \vee \neg P_n(\bar{x})  \vee A_i \geq c_i
\vee A_j < c_j \vee A_k < c_k \vee A_k > c_k \vee A_l = c_l \vee
\ldots) \label{eq:ICvee1}
\end{equation}
This formula shows that since the repairs are done by attributes
updates, the only way we have of solving an inconsistency is by
fixing at least one of the values of a fixable attribute. Let
$\nit{ic}$ be a constraint in $\IC$ such that ${\cal
I}(D,\nit{ic},t) \neq \emptyset$ and $I$ be a violation set $I \in
{\cal I}(D,\nit{ic},t)$. Now, since $\nit{ic} \in \IC$, $\nit{ic}$
is a consistent constraints. Then for each fixable attribute $A$
in $\nit{ic}$ we are able to derive an interval $[c_l,c_u]$ such
that if the value of $A$ is in it, we would restore the
consistency of $I$. For example if we have a constraint in form of
equation (\ref{eq:ICvee1}) with $A \leq 5$, then, if we want to
restore consistency by modifying $A$ we would need to have $A \in
~(-\infty,5]$. If the constraint  had also $A \geq 1$ the interval
would be $[1,5]$. Since $t$ has at least one fixable attribute and
each fixable attribute has an interval, it is always possible to
adjust the value of that fixable attribute to a value in the
interval $[c_l,c_u]$ and restore consistency. By finding the
adjustment that minimizes the distance from the
original tuple we have find a local fix  for the tuple $t$.}\\

The {\em borders} of an attribute in an extended linear denial
correspond to the surfaces of the semi-spaces determined by the
built-in atoms in it.

\begin{propositionA} \label{prop:prelimits} Given a database $D$ and a set of
linear denials $\IC$, where equalities and joins can only exist
between rigid attributes, the values in every fixable attributes
in a local fix $t'$ (c.f. Definition \ref{def:locfix}) of a tuple
$t \in D$ will correspond to the original value in $t$ or to a
border of a constraint in $\IC$. Furthermore, the values in every
attributes of a tuple $t' \in D'$ will correspond to the original
value of the attribute in the tuple in $D$ or to a border of a
constraint in $\IC$. \boxtheorem
\end{propositionA}

\vspace{2mm}\dproof{}{First we will replace in all the constraints
$X \leq c$ by $X < (c+1)$, $X\geq c$ by $X
> (c-1)$ and $X = c$ by $(X > (c-1) \wedge X < (c+1))$. We can do
this because we are dealing with integer values. Then, a
constraint $\nit{ic}$ would have the form $\forall \bar{x}
\neg(P_1(\bar{x}), \ldots,P_n(\bar{x}),$ $A_i < c_i, A_j > c_j,
A_k \neq c_k ,\ldots)$ and can be rewritten as \vspace{-.3cm}
\begin{equation} \forall \bar{x} (\neg P_1(\bar{x}) \vee \ldots \vee \neg
P_n(\bar{x}) \vee A_i \geq c_i  \vee A_j \leq c_j \vee A_k = c_k
\vee \ldots) \label{eq:ICvee}
\end{equation}

\vspace{-.3cm} \noindent This formula shows that since the repairs
are done by attributes updates, the only way we have of solving an
inconsistency is by fixing at least one of the values of a fixable
attribute. This would imply to change the value of a fixable
attribute $A_i$ to something equal or greater than $c_i$,
 to change the value of a fixable attribute $A_j$ to a value
equal or smaller  than $c_j$ or to  change the value of attribute
$A_k$ to $c_k$.

If $D$ is consistent wrt $\IC$ then there is a unique LS-fix
$D'=D$ and all the values are the same as the original ones and
therefore the proposition holds. If $D$ is inconsistent wrt $\IC$
then there exists a tuple $t$ with at least one fixable attribute
and a set $\IC_{t} \subseteq \IC$ such that for every $\nit{ic}
\in \IC_t$ it holds ${\cal I}(D,\nit{ic},t) \neq \emptyset$. If
$\IC_t$ is an inconsistent set of constraints then there exists no
local fix and the proposition holds. If $\IC_t$ is consistent but
there is at least one constraint with no fixable attributes
involved then, since it is not possible to modify any attribute in
order to satisfy the constraint, there is no local fix and the
proposition holds.

So we are only missing to prove the proposition for $\IC_t$
consistent and with at least one fixable attributes for each
$\nit{ic}$ in $\IC_t$. From Lemma A.\ref{lemma:existLF} we know
that there exists a local fix for $t$. Also, since  $\IC_t$ is
consistent, using the same arguments as in proof of Lemma
A.\ref{lemma:existLF}, it is possible to define for each fixable
attribute $A$ an interval such that if the value of $A$ is in it
we would restore the consistency of the violation sets for
constraints in $\IC_t$ involving $t$. Then, we need to prove that
if a value of an attribute, say $A$, of a local fix $t'$ of $t$ is
different than the one in $t$, then the value corresponds to one
of the closed limits of the interval for $A$. Let us assume that
an attribute $A$ is restricted by the constraints to an interval
$[c_l,c_u]$ and that the local fix $t'$ takes for attribute $A$ a
value strictly smaller than $c_u$ and strictly greater than $c_l$.
Without lost of generality we will assume that the value of
attribute $A$ in $t$ is bigger than $c_u$. Let $t''$ be a tuple
with the same values as $t'$ except that the attribute $A$ is set
to $c_u$. $t''$ will have the same values in the rigid attributes
as $t$ and also $S(t,t')=S(t,t'')$ since the value of $A$ in $t''$
is still in the interval. We also have that $\Delta(\{t\},
\{t''\}) \leq \Delta(\{t\}, \{t'\})$. This implies that $t'$ is
not a local fix and we have reached a contradiction.

For the second part of the proposition, the proof of the first
part can be easily extended to prove that the values in $D'$ will
correspond to a border of a constraint in $\IC$, because the
LS-fixes are combination of local fixes.}

\vspace{.3cm}\defproof{Theorem \ref{theo:exist}}{Hilbert's 10th
problem on existence of integer solutions to diophantine equations
can be reduced to our problem. Given a diophantine equation, it is
possible to construct a database $D$ and a set of ICs $\IC$ such
that the existence of an LS-fix for $D$ wrt $\IC$ implies the
existence of a solution to the equation, and viceversa. An example
can be found in Appendix \ref{sec:hilbert}.}

\vspace{.3cm}
\defproof{Proposition \ref{lemma:reduc}}{First for the skeptical semantics.
Given a database instance $D$, consider the instance $(D,
\nit{no})$ for $\nit{CQA}(\nit{False},\IC, \nit{Sk})$,
corresponding to the question ``Is there an LS-fix of $D$ wrt
$\IC$ that does not satisfy ${\it False}$?" has answer ${\it Yes}$
iff the class of LS-fixes of $D$ is empty. For the majority
semantics, for the instance $(D,\it{no})$ for
$\nit{CQA}(\nit{True},\IC, \nit{Maj})$, corresponding to the
question ``Is it not the case that the majority of the LS-fixes
satisfy $\nit{True}$?", we get answer ${\it yes}$ iff the set of
LS-fixes is empty.}

\vspace{.3cm}\defproof{Theorem \ref{lemma:squa-member}}{(a) First
of all, we notice that a linear denial with implicit equalities,
i.e. occurrences of a same variable in two different database
atoms, e.g. $\forall X,Y,Z \neg (R(X,Y),Q(Y,Z),$ $Z
> 3)$, can be replaced by its {\em explicit version} with explicit
equalities, e.g. $\forall X,Y,Z,W \neg (R(X,Y),$ $Q(W,Z), Y=W, Z >
3)$.

Let $n$ be the number of tuples in the database, and $l$ be the
number of attributes which participate in $\IC$. They are those
that appear in built-in predicates in the explicit versions of the
ICs that do not belong to a key or are equal to a key (because
they are not allowed to change). For example, given the denial
$\neg(P(X,Y),Q(X,Z),Y>2)$, since its explicit version is
$\neg(P(X,Y),Q(W,Z),Y>2,X=W)$, the number $l$ is 1 (for $Y$) if
$X$ is a key for $P$ or $Q$, and $3$ if $X$ is not a key (for $Y,
X, W$).

If there exist an LS-fix $D'$ with $\Delta(D,D') \leq k$, then no
value in a fixable attribute in $D'$ differs from its
corresponding value (through the key value) in $D$ by more than
$\sqrt k$. In consequence, the size of an LS-fix may not differ
from the original instance by more than $l \times n \times
\nit{bin}(k)/2$, where $\nit{bin}(k)$ is the size of the binary
representation of $k$. Thus, the size of an LS-fix is polynomially
bounded by the size of $D$ and $k$. Since we can determine in
polynomial time if $D'$ satisfies the ICs and if the distance is
smaller than $k$, we obtain the result.\\
\\
(b) Membership: According to Proposition \ref{lemma:existence},
there is an LS-fix at a square distance $\leq k$ iff there is an
instance $D'$ with the same key values that satisfies \IC\ at a
square distance $\leq k$. We use Proposition
\ref{lemma:squa-member}.\\
\\
Hardness: We can reduce Vertex Cover (VC) to $\nit{DFP}(\IC_0)$
for a fixed set of denials $\IC_0$. Given an instance
$(\cal{V},\cal{E}),$ $k$ for VC, consider a database $D$ with a
relation $E(X,Y)$ and key $\{X,Y\}$ for the edges of the graph,
and a relation for the vertices $V(X,\nit{Chosen})$, where $X$ is
the key  and attribute $\nit{Chosen}$, the only fixable attribute,
is initially set to $0$. The constraint $\IC:$ $\forall X,Y,C_1,
C_2 \neg (E(X,Y) \land V(X, C_1) \land V(Y, C_2) \land C_1 < 1
\land C_2 < 1)$ expresses that for any edge, at least one of the
incident vertices is covered. A vertex cover of size $k$ exists
iff there exists an LS-fix of $D$ wrt $\IC$ at a distance $\leq
k$. The encoding is polynomial in the size of the original graph.}

\vspace{.3cm}\defproof{Theorem \ref{cor:finNPcomp}}{(a) For
hardness, linear denials are good enough. We reduce the graph
3-colorability problem to $\nit{NE}(\IC_0)$, for a fixed set
$\IC_0$ of ICs. Let $\mathcal{G} = (\mathcal{V}, \mathcal{E})$ be
an undirected graph with set of vertexes $\mathcal{V}$ and set of
edges $\mathcal{E}$. Consider the following database schema,
instance $D$, and set $\IC_0$ of ICs:

1. Relation $Vertex(\nit{Id, Red, Green, Blue})$ with key
$\nit{Id}$ and domain $\mathbb{N}$ for
  the last three attributes, actually the only three fixable attributes in the
database; they can be
  subject to changes. For each $v \in \mathcal{V}$ we have the tuple
  $(v, 0, 0, 0)$ in $\nit{Vertex}$ (and nothing else).

  2. Relation $Edge(id_1, id_2)$; and for each $e = (v_1, v_2) \in \mathcal{E}$,
there are the tuples
  $(v_1, v_2), (v_2, v_1)$ in $Edge$. This relation is not
  subject to any fix.

 3. Relation $\nit{Tester(Red,Green, Blue})$, with extension $(1,0,0), (0,1,$
$0),
 (0,0,1)$. This relation is not subject to any fix.

4. Integrity constraints:\\
 $\forall ixyz \neg (\nit{Vertex}(i,x,y,z), x <1, y < 1, z <
 1)$;
 $\forall ixyz \neg (\nit{Vertex}(i,x,y,z), x > 1)$ (the same
 for
$y, z$);
 $\forall ixyz \neg (\nit{Vertex}(i,x,y,z), x  = 1, y = 1, z =
 1)$;
$\forall ixyz \neg (\nit{Vertex}($ $i,x,y,z), x = 1, y = 1)$;
etc.\\
$\forall i j x y z\neg (\nit{Vertex}(i, x, y, z),
\nit{Vertex}(j,x,y,z),\nit{Edge}(i,j), \nit{Tester}(x,y,z)$.

The graph is 3-colorable iff the database has an LS-fix wrt
$\IC_0$. The reduction is polynomial in the size of the graph. If
there is an LS-fix of the generated instance, then the graph is
3-colorable. If the graph is colorable, then there is a consistent
instance with the same key values as the original instance; then,
by Proposition \ref{lemma:existence}, there is an LS-fix.

For membership, it suffices to prove that if an LS-fix exists,
then its square distance to $D$ is polynomially bounded by the
size of $D$, considering both the number of tuples and the values
taken by the fixable attributes.

We will show that if an LS-fix $D'$ exists, then all the values in
its fixable attributes  are bounded above by the maximum of $n_1 +
n +1$ and $n_2 + n +1$, where $n$ is the number of tuples in the
database, $n_1$ is the maximum absolute value in a fixable
attribute in $D$, and $n_2$ is the maximum absolute value of a
constant appearing in the ICs.

The set of denial ICs put in disjunctive form gives us a
representation for all the ways we have to restore the consistency
of the database. So, we have a constraint of the form~ $\varphi_1
\land \varphi_2 \cdots \varphi_m$, where each $\varphi_i$ is a
disjunction of negated database atoms and inequalities, e.g.
something like $\neg P(X,Y,Z) \lor \neg R(X_1, Y_1) \lor X \leq
c_1 \lor Y \leq c_2 \lor Z \neq Y_1$. Since fixes can be obtained
by changing values of non key attributes, each tuple in a fix is
determined by a set of constraints, each of which is a disjunction
of atoms of the form $X_i \theta_i c_m$ or $X_i \neq Y_j$, where
$\theta_i$ is an inequality of the form $\leq, \geq, <, >$. E.g.
from  $\neg P(X,Y,Z) \lor \neg R(X_1, Y_1) \lor X \leq c_1 \lor Y
\leq c_2 \lor Z \neq Y_1$ we get $X \leq c_1 \lor Y \leq c_2 \lor
Z \neq Y_1$, which for a specific tuple becomes $Y \leq c_2 \lor Z
\neq Y_1$ if  $X$ is part of the key and its specific value for
the tuple at hand does not satisfy $X \leq c_1$ (otherwise we drop
the constraint for that tuple). In any case, every tuple in a fix
can take values in a space $S$ that is the intersection of the
half-spaces defined by inequalities of the form $X_i \theta_i c_m$
minus the set of points determined by the non-equalities $X_i \neq
Y_j$.

If there is a set of values that satisfies the resulting
constraints, i.e. if there is an instance with the same key values
that satisfies the ICs, then we can find an LS-fix at the right
distance: if the difference between any value and $max(c_1,
\cdots, c_l)$ is more than $n + 1$ (the most we need to be sure
the inequalities $X_i \neq Y_j$ are satisfied), then we
systematically change values by 1, making them closer to the
borders of the half-spaces, but still keeping the points within
$S$.

\noindent (b) $\nit{coNP}$-hardness follows from Proposition
\ref{lemma:reduc} and part (a).}

\vspace{.3cm}\defproof{Theorem \ref{theo:checking}}{(a) We reduce
3-SAT's complement to LS-fix checking for a fixed schema and set
of denials $\IC$. We have a table $\nit{Lit}(l,\bar{l})$ storing
complementary literals (only), e.g. $(p,\neg p)$ if $p$ is one of
the variables in the instance for SAT. Also a table $\nit{Cl}$
storing tuples of the form $(\varphi, l, k)$, where $\varphi$  is
a clause (we assume all the clauses have exactly 3 literals, which
can be simulated by adding extra literals with unchangeable value
$0$ if necessary), $l$ is a literal in the clause, and $k$ takes
value $0$ or $1$ (the truth value of $l$ in $\varphi$). The first
two arguments are the key of $C$. Finally, we have a table
$\nit{Aux}(K,N)$, with key $K$ and fixable numerical attribute
$N$, and a table $\nit{Num}(N)$ with a rigid numerical attribute
$N$.

Given an instance $\Phi = \varphi_1 \wedge \cdots \wedge
\varphi_m$ for 3-SAT, we produce an initial extension $D$ for the
relations in the obvious manner, assigning arbitrary truth values
to the literals, but making sure that the same literal takes the
same truth value in every clause, and complementary literals take
complementary truth values. $\nit{Aux}$ contains $(0,0)$ as its
only tuple; and $\nit{Num}$ contains $(s +1)$, where $s$ is the
number of different propositional variables in $\Phi$.

Consider now the following set of denials:\\
(a) $\neg (\nit{Cl}(\varphi,L,U), U > 1)$; $\neg
(\nit{Cl}(\varphi,L,U), U <~0)$ ~~(possible truth values).\\ (b)
$\neg(\nit{Cl}(\varphi,L,U),$ $\nit{Cl}(\psi,L,$ $V),$ $U \neq V)$
~~(same value for a literal everywhere).\\ (c)
$\neg(\nit{Cl}(\varphi,L,U),$ $\nit{Cl}(\psi,L',V),
\nit{Lit}(L,L'), U = V)$ ~~(complementary
literals).\\
(d) $\neg (\nit{Cl}(\varphi,L,U),
\nit{Cl}(\varphi,L',V),\nit{Cl}(\varphi,L'',W), U = V = W = 0, L
\neq L', ...,$ $\nit{Aux}(K,N), N = 0)$ (each clause becomes true).\\
(e) $\neg (\nit{Num}(Z), \nit{Aux}(K,N), N \neq 0, N \neq Z)$
~~(possible values).

It holds that the formula is unsatisfiable iff the instance $D'$
that coincides with $D$ except for $\nit{Aux}$ that now has the
only tuple $(0,s+1)$ is an LS-fix of $D$ wrt $\IC$. Thus, checking
$D'$ for LS-fix is enough to check unsatisfiability.

For membership to $\nit{coNP}$, for an initial instance $D$,
instances $D'$ in the complement of $\nit{Fix}(\IC)$ have
witnesses $D''$ that can be checked in polynomial time, namely
instances $D''$ that have the same key values as $D$, satisfy the
ICs, but $\Delta(D,D'')< \Delta(D,D')$.

\noindent (b) For the first claim on CQA, let $\IC$ and a query
$Q$ be given. The complement of CQA is in $\nit{NP}^{\nit{coNP}}$:
Given an instance $D$, non deterministically choose an instance
$D'$ with $D' \not \models Q$ and $D'$ a fix of $D$. The latter
test can be done in $\nit{coNP}$ (by part (a)). But
$\nit{NP}^{\nit{coNP}} = NP^{\Sigma_1^P} = \Sigma_2^P$. In
consequence, CQA belongs to $\nit{co}\Sigma_2^P = \Pi^P_2$.

For the second claim, we prove hardness of {\em CQA} by a
$\nit{LOGSPACE}$-reduction from the following problem \cite[Theo.
3.4]{krentel}: Given a Boolean formula $\psi(X_1, \cdots, X_n)$ in
3CNF, decide if the last variable $X_n$ is equal to $1$ in the
lexicographically maximum satisfying assignment (the answer is
false if $\psi$ is not satisfiable).

Given the clauses $C_1, \ldots, C_m$ in $\psi$, we create a
database $D$ with relations $\nit{Var(id, tr, fa, weight)},
\nit{Cl(id, var_1, val_1, var_2, val_2, var_3, val_3)}$ and
constraints:
\begin{enumerate}
\item $\forall id, tr, fa \neg(Var(id, tr, fa, \_) \land tr \leq 0
\land fa \leq 0)$
 \item  $\forall id, tr, fa \neg(Var(id, tr, fa,
\_) \land tr \geq 1 \land fa \geq 1)$
\item $\forall id, tr, fa, w
\neg(Var(id, tr, fa, w) \land fa = 1 \land w > 0)$
\item  $\forall
id, v_1, x_1, v_2, x_2, v_3, x_3  \neg (Cl(id, v_1, x_1, v_2, x_2,
v_3, x_3)
 \land Var(\_, v_1, x'_1)  \land Var(\_, v_2, x'_2)$
  $\land Var(\_, v_3, x'_3) \land x_1 \not = x'_1  \land x_2 \not = x'_2 \land v_3 \not =
  x'_3)$
\end{enumerate}
The extended denial constraint in 4. could be replaced by eight
non-extended denial constraints.

 For each variable $X_i$, insert a
tuple $(X_i, 0, 0, 2^{n-i})$ into $\nit{Var}$. In binary encoding,
the values $2^{n-i}$ are polynomial in the size of original
formula. For each clause $C_i = l_{i_1} \lor l_{i_2} \lor
l_{i_3}$, insert a tuple $(C_i, X_{i_1}, \tilde{l}_{i_1}, X_{i_2},
\tilde{l}_{i_2}, X_{i_3}, \tilde{l}_{i_3})$ into $\nit{Cl}$, where
$\tilde{l}_{i_j}$ is equal to $1$ in case of positive occurrence
of variable $X_{i_j}$ in $C_i$ and equal to $0$ for negative
occurrence. For example, for $C_6 = X_6 \lor \neg X_9 \lor
X_{12}$, we insert $(C_6, X_6,1, X_9, 0,  X_{12}, 1)$.

It is easy to see that  all the fixes $D$ represent satisfying
assignments for $\psi$, such that in case a tuple $(X_i, 1, 0,\_)$
is in a fix, then value $\nit{true}$ must be assigned to variable
$X_i$; and value $\nit{false}$ must be assigned to  $X_i$ in case
 $(X_i, 0, 1,\_)$ is in the fix. If $\psi$ is unsatisfiable, then
there are no fixes.

Let us now consider the cost of a repair. Assume  that $S_1 =
x_{i^1_1}, \ldots, x_{i^1_{m_1}}$ and $S_2 = x_{i^2_1}, \ldots,
x_{i^2_{m_2}}$ are satisfying assignments with $S_1 \prec S_2$
under lexicographical order. Since $S_1 \prec S_2$, there exists
an integer
 $m$ such that $i^1_m < i^2_m$, while for all $j < m$, $i^1_j =
 i^2_j$, by definition of lexicographical order. The cost of repair
 $S$ is equal to
 $2^{n - {i_1}} + 2^{n_{i_2}} +\cdots +
 2^{n_{i_{m_X}}} + n$, because (a) we have to update attribute $\nit{weight}$ for each variable
 that is
 assigned value $\nit{true}$, and (b) for each tuple in relation $\nit{Var}$, attribute $\nit{tr}$ or attribute
 $\nit{fa}$ is changed by $1$.

 Because of (a), there exists a term $2^{n-i^1_m}$
 in the cost of $S_1$ that is bigger than the sum of all terms
 in $S_2$, with index  $\geq i^2_m$.
 So, the cost of the fix representing $S_1$ is greater than the cost
 of the fix representing $S_2$. In consequence, the minimal fix
 would be the maximum according to the lexicographical order of satisfying
 assignments.

 The answer to the ground atomic
query $\nit{Var}(X_n, 1, 0, 1)$ is {\em true} iff
 $X_n$ takes the value $1$ in the l.o. greatest assignment.}

 \vspace{3mm}
\defproof{Theorem \ref{prop:minCover}}{The reduction can be
established with a fixed set $\IC_0$ of ICs. Given an undirected
graph ${\cal G} = ({\cal V}, {\cal E})$, consider a database with
relations $V(X, Z), E(U,W)$, where $X$ is a key and $Z$ is the
only fixable attribute and takes values in $\{0,1\}$ (which can be
enforced by means of the linear denials $\forall X \forall
Z\neg(V(X,Z), Z
>1)$, $\forall X \forall Z\neg(V(X,Z), Z < 0)$ in  $\IC_0$). Intuitively, $Z$
indicates with $1$ if the vertex $X$ is in the cover, and with $0$
otherwise. Attributes $U,V$ are vertices and then, non numerical.

In the original database $D$ we have the tuples $V(e,0)$, with $e
\in {\cal V}$; and also the tuples $E(e_1,$ $e_2)$ for $(e_1,$
$e_2) \in {\cal E}$. Given the linear constraint~ $$\forall X_1
Z_1 X_2 Z_2 \neg (V(X_1, Z_1), V(X_2, Z_2),E(X_1, X_2), Z_1 = 0,
Z_2 = 0)$$ in $\IC_0$, the LS-fixes of the database are in
one-to-one correspondence with the vertex covers of minimal
cardinality.

For the query $Q^{(k)}\!: q(sum(Z)), sum(Z) <k$, with $q(sum(Z))
\leftarrow V(X,Z)$, the instance $(D,\nit{yes})$ for consistent
query answering  under brave semantics has answer {\em No}, (i.e.
$Q^{(k)}$ is false in all LS-fixes) only for every $k$ smaller
than the minimum cardinality $c$ of a vertex cover. }

In the previous proof, the first two ICs in $\nit{IC}_0$ can be
eliminated, and the third one is local (c.f. Definition
\ref{def:local}). In consequence, the theorem also holds for local
denials.

\vspace{.3cm}\defproof{Proposition \ref{lem:dfp-snphard}}{By
reduction from the $\nit{MAXSNP}$-hard problem \emph{B-Minimum
Vertex Cover} (BMVC), which asks to find a minimum vertex cover in
a graph whose nodes have a bounded degree \cite[chap.
10]{hochbaum}. We start by encoding the graph as in the proof of
Theorem \ref{prop:minCover}. We also use the same initial database
$D$. Every LS-fix $D'$ of $D$ corresponds to a minimum vertex
cover ${\cal V}'$ for ${\cal G}$ and vice versa, and it holds
$|{\cal V}'| = \Delta(D,D')$. This gives us an $L$-reduction from
BMVC to $\nit{DFP}$ \cite{papadimitriou94}.}

\begin{lemmaA} \label{lemma:exLS} Given a database $D$ and a set of
consistent local denials $\IC$, there will always exist an LS-fix
$D'$ of $D$ wrt $\IC$. \boxtheorem
\end{lemmaA}

\dproof{Lemma A.\ref{lemma:exLS}}{ As shown in proof of Lemma
A.\ref{lemma:existLF} for every fixable attribute  in $F$ it is
possible to define, using the integrity constraints in $\IC$, an
interval $[c_l,c_u]$ such that if the value of attribute $A$ is in
that interval there is no constraint $\nit{ic} \in \IC$ with a
built-in involving $A$ such that ${\cal I}(D,\nit{ic},t) \neq
\emptyset$. Let $D''$ be a database constructed in the following
way: for every tuple $t \in D$  such that the value of a fixable
attribute does not belong to its interval,  replace its value by
any value in the interval. Clearly $D'$ will be a fix but will not
necessarily be an LS-fix. By Proposition \ref{lemma:existence} we
know there exists an LS-fix $D'$ for $D$ wrt $\IC$. }

\begin{definition} \label{def:newvio} \em Given a database $D$ and a set of ICs
$\IC$, a local fix $t'$ for a tuple $t$ {\em does not} {\em
generate new} {\em violations} if  $\bigcup_{\nit{ic} \in \IC}
(\bigcup_{l \in D'} {\cal I}(D',\nit{ic},l)$ $\smallsetminus$ $
\bigcup_{l \in D} {\cal I}(D,\nit{ic},l)) = \emptyset$~ for
~$D'=(D \smallsetminus \{t\}) \cup \{t'\}$. \boxtheorem
\end{definition}

\begin{lemmaA}\label{lem:ncicp}
For a set $\IC$ of local denials,  if $t'$ is a local fix of a
tuple $t$, then $t'$ does not generate new
violations\footnote{c.f. Definition \ref{def:newvio}} in database
$D$ wrt $\IC$. Furthermore, this holds also for $t'$ a ``relaxed"
local fix where the distance to $t$ is not necessarily minimal
\boxtheorem
\end{lemmaA}
\dproof{Lemma A.\ref{lem:ncicp}}{Tuple $t'$ can only differ from
$t$ in the value of fixable attributes. Let us assume that one of
the modified values was for an attribute $A$. Since we have local
constraints, attribute $A$ can only be in the constraints related
either to  $<$ and $\leq$ or to $>$ and $\geq$, but not both.
Without lost of generality, we will assume that the constraint is
written as in equation \ref{eq:ICvee1} and that $A$ is related
only to $>$ and $\geq$. Since $t'$ is a local fix, $S(t,t')$ is
not empty and there is a set $\IC_t$ of constraints for which $t'$
solves the inconsistency in which $t'$ was involved. There is an
interval $[c_l,+\infty)$ for $A$ that can be obtained by the
limits given in $\IC_t$ that show the values of $A$ that would
force the satisfaction of the constraints in $\IC_t$ that have
attribute $A$ in an inequality. This shows that the value of
attribute $A$ in $t'$ is bigger than the value of $A$ in $t$.

For $D'=(D \smallsetminus \{t\}) \cup \{t'\}$ we need to prove
that $\bigcup_{\nit{ic} \in \IC} (\bigcup_{l \in D'} {\cal
I}(D',\nit{ic},l) ~\smallsetminus~$ $ \bigcup_{l \in D} {\cal
I}(D,\nit{ic},l)) = \emptyset$. By contradiction let us assume
that for a constraint $\nit{ic} \in \IC$ there exists a violation
set $I$ such that $I \in \bigcup_{l \in D'}{\cal
I}(D',\nit{ic},l)$ and $I \not \in \bigcup_{l \in D}{\cal
I}(D,\nit{ic},l)$. There are two cases to consider (with
$(I,\nit{ic}$, we indicate that $I$ is a violation set for IC
$\nit{ic}$): \vspace{-.15cm}\begin{itemize} \item $(I,\nit{ic})
\in S(t,t')$. Then $I \in {\cal
   I}(D,ic,t)$, but since we wanted an $I \not \in
\bigcup_{l \in D}{\cal I}(D,$ $\nit{ic},l)$ this is not possible.
\item $(I,\nit{ic}) \not \in S(t,t')$. Then we have two
   possibilities $I \not \in {\cal I}(D,ic,t)$  or $((I \smallsetminus
\{t\}) \cup \{t'\})  \not \models ic$.
\begin{itemize}
\item Let us consider first that $I \not \in {\cal I}(D,ic,t)$. We
have that $I \in \bigcup_{l \in D'}{\cal I}(D',$ $\nit{ic},l)$ and
since $t'$ is the only difference between $D$ and $D'$ we have $I
\in {\cal I}(D',\nit{ic},t')$. Since all the constraints can only
have attribute $A$ with $>$ or $\geq$ we now that in particular
$\nit{ic}$ does. Since $I \not \in {\cal I}(D,ic,t)$ we know that
$A$ satisfied the condition in $\nit{ic}$ and since we know that
$t'$ has a bigger value than in $t$, it is not possible to
generate an inconsistency in $D'$. We have reached a
contradiction. \item Let us consider $((I \smallsetminus \{t\})
\cup \{t'\}) \not \models ic$. Then $I \in {\cal I}(D',ic,t')$.
From our assumption
 $I \not \in \bigcup_{l \in D}{\cal
I}(D,\nit{ic},l)$.This is the same situation analyzed in previous
item.
\end{itemize}
\end{itemize}
\vspace{-.3cm} In all the cases we have reached contradiction and
therefore the proposition is proved. Since we never used the
property of minimal distance between $t'$ and $t$, the second part
of the Lemma is also proved.}

\begin{propositionA} \label{prop:itworks}For local denials it always exists
an LS-fix for a database $D$; and for every LS-fix $D'$, $D'
\smallsetminus D$ is a set of local fixes. Furthermore, for each
violation set $(I,\nit{ic})$, there is a tuple $t \in I$ and a
local fix $t'$ for $t$, such that $(I,\nit{ic}) \in S(t,t')$.
\boxtheorem
\end{propositionA}

\dproof{Proposition A.\ref{prop:itworks}}{ Since each attribute A
can only be associated to $<$  or $>$ built-ins, but not both, it
is clear that set of local denials is always consistent. By Lemma
A.\ref{lemma:exLS}, there always exists an LS-fix $D'$. Now we
need to prove that $D' \smallsetminus D$ is a set of local fixes.
By contradiction assume that $t' \in (D' \smallsetminus D)$ is not
a local fix of the tuple $t$. This can happen in the following
situations: \begin{itemize}
    \item $t$ was consistent. From Lemma A.\ref{lem:ncicp}
    we know that no new inconsistencies can be added by the
    modifications done to the other tuples and therefore $t$ is not related
    to any inconsistency. Then $D^\star=D' \smallsetminus \{t'\} \cup
    \{t\}$ is also consistent and $\Delta(D,D^\star) <
    \Delta(D,D')$. But $D'$ is an LS-fix so this is not possible.
    \item $t$ is involved at least in one violation set. If $S(t,t')=
    \emptyset$ then $t'$ is not solving any violation set and
    therefore $D^\star=D' \smallsetminus \{t'\} \cup
    \{t\}$ is also consistent and $\Delta(D,D^\star) <
    \Delta(D,D')$. But $D'$ is an LS-fix so this is not possible.
    Now, if $S(t,t')\neq\emptyset$, from Lemma A.\ref{lemma:exLS}, considering
    $D=\{t\}$ and $\IC= \{\nit{ic} | (I,\nit{ic}) \in S(t,t')\}$,
    there exists an LS-fix $D'$ of $D$, i.e. there exists a local fix $t''$ such that
    $S(t,t'')=S(t,t')$. Since $t''$ is a local fix we know that $\Delta(\{t\},\{t''\}) \leq
    \Delta(\{t\},\{t'\})$. They cannot be equal that would imply that  $t'$ is
    a local fix and it is not. Then $D^\star=D' \smallsetminus \{t'\} \cup
    \{t\}$ is also consistent and $\Delta(D,D^\star) <
    \Delta(D,D')$. Again, this is not possible because $D'$ is an
    LS-fix
\end{itemize}

\vspace{-.3cm} \noindent The second part of the proposition can be
proved using Lemma A.\ref{lemma:exLS} and considering a database
$D=I$ and a set of constraints $\IC=\{\nit{ic}\}$.}\\

\defproof{Proposition \ref{prop:still}}{For the first claim, membership follows from
Theorem \ref{lemma:squa-member}(b); and for hardness, we can do
the same reduction as in Theorem \ref{lemma:squa-member}(b),
because the ICs used there are local denials. For the second
claim, it is not difficult to see that the non-local denials in
the proof of Proposition \ref{lem:dfp-snphard} can be eliminated.}

\begin{propositionA}\label{prop:LFcombined} For a database $D$
and a set of local denial constraints $\IC$:
\begin{enumerate}
\item For a set of local fixes $\{t_1, \dots t_n\}$ of a tuple $t$
there always exists a local fix $t^\star$ such that $S(t,t^\star)=
\bigcup_{i=1}^n S(t,t_i)$. \item For local fixes $t'$, $t''$ and
$t'''$ of a tuple $t$ with $S(t,t''')=S(t,t') \cup S(t,t'')$, it
holds that $\Delta(\{t\},\{t'''\}) \leq
\Delta(\{t\},\{t'\})+\Delta(\{t\},\{t''\})$. \boxtheorem
\end{enumerate}
\end{propositionA}
\dproof{Proposition A.\ref{prop:LFcombined}}{First we prove item
(1). Let $IC_t=\{\nit{ic}|{\cal I}(D,\nit{ic},t) \neq \emptyset$
and $IC_S(t,t')=\{\nit{ic}|(I,\nit{ic})\in S(t,t')\}$. From Lemma
A.\ref{lemma:exLS}, considering $D=\{t\}$ and $\IC$ any subset of
$\IC_t$, there always exists an LS-fix $D'$ of $D$. This LS-fix is
a local fix of tuple $t$ with $IC_S(t,t')=\IC$. Since we can find
a local fix for any $\IC \subseteq S_t$ then clearly the lemma can
be satisfied.

Now we will prove item (2).  If the fixable attributes that where
modified in $t'$ and $t''$ are disjoint, then $t'''$ when
combining the modifications I'll get $\Delta(\{t\},\{t'''\})=
\Delta(\{t\},\{t'\})+\Delta(\{t\},\{t''\})$. Now, we will consider
the case were $t'$ and $t''$ have at least one fixable attribute,
say $A$ that is modified by both local fixes. In this case $t'''$
will have a value in $A$ that solves the inconsistencies solved by
and $t'$ and $t''$. This value will in fact correspond to the
value of $A$ in $t'$ or $t''$ and therefore we will have that
$\Delta(\{t\},\{t'''\})<
\Delta(\{t\},\{t'\})+\Delta(\{t\},\{t''\})$. Let $M$ be the set of
attributes that are modified both by $t'$ and $t''$, we can
express the relation as follows: $\Delta(\{t\}, \{t'''\}) =
\sum_{A \in {\cal F}}
   (\pi_{\!A}(t)-
   \pi_{\!A}(t'''))^2$ $= \sum_{A \in {\cal F}}
   (\pi_{\!A}(t)-
   \pi_{\!A}(t'))^2$ $+\sum_{A \in {\cal F}}
   (\pi_{\!A}(t)-
   \pi_{\!A}(t''))^2$ $- \sum_{A \in {\cal M}}
   Min\{(\pi_{\!A}(t)-
   \pi_{\!A}(t'))^2,(\pi_{\!A}(t)-
   \pi_{\!A}(t''))^2 \}$}

\begin{propositionA} \label{prop:LScover} If an optimal cover ${\cal C}$ for the instance $(U,{\cal S})$ of
$MWSCP$ has more than one $S(t,t')$ for a tuple $t$, then there
exists another optimal cover ${\cal C'}$ for $(U,{\cal S})$ with
the same total weight as ${\cal C}$ but with only one $t'$ such
that $S(t,t')\in {\cal C}$. Furthermore, $D({\cal C})$ is an
LS-fix of $D$ wrt $\IC$ with $\Delta(D,D(\cal C))$ equal to the
total weight of the cover $\cal C$. \boxtheorem
\end{propositionA}

\dproof{Proposition A.\ref{prop:LScover}}{To prove the first part,
let us assume that  $S(t,t'), S(t,t'') \in {\cal C}$. From
Proposition A.\ref{prop:LFcombined} there exists an $S(t,t''') \in
{\cal S}$ such that $S(t,t''')=S(t,t') \cup S(t,t'')$, i.e such
that it covers the same elements as $S(t,t')$ and $ S(t,t'')$.
>From Proposition A.\ref{prop:LFcombined} $\Delta(\{t\},\{t'''\})
\leq \Delta(\{t\},\{t'\})+\Delta(\{t\},\{t''\})$ and therefore
that weight of $S(t,t''')$ is smaller or equal than the sum of the
weight of the original two sets. If $\Delta(\{t\},\{t'''\}) <
\Delta(\{t\},\{t'\})+\Delta(\{t\},\{t''\})$ we would have that
${\cal C}$ is not an optimal solution so this is not possible.
Then  $\Delta(\{t\},\{t'''\}) =
\Delta(\{t\},\{t'\})+\Delta(\{t\},\{t''\})$. Then, if we define
${\cal C'} = ({\cal C} \smallsetminus \{S(t,t'), S(t,t'')\}) \cup
\{S(t,t''')\}$ we will cover all the elements and we will have the
same optimal weight.

Now we need to prove that given $D({\cal C})$ is an LS-fix.
$D({\cal C})$ is obtained by first calculating ${\cal C}'$ and
therefore we have an optimal cover with at most one $S(t,t')$ for
each tuple $t$. Then $D({\cal C})$ is obtained by replacing $t$ by
$t'$ for each $S(t,t') \in {\cal C}$.  It is direct that $D({\cal
C})$ has the same schema as $D$ and that it satisfies the key
constraints. Now, since $\cal C'$ covers all the elements, all the
inconsistencies in $D$ are solved in $D({\cal C})$. From Lemma
A.\ref{lem:ncicp} the local fixes $t'$ do not add new violations
and therefore $D({\cal C}) \models \IC$ and $D({\cal C})$ is a
fix. We are only missing to prove that  $D({\cal C})$ minimizes
the distance from $D$. Clearly $\Delta(D,D({\cal C})) = \sum_{t
\in D} \Delta(\{t\},\{t'\})$ $= \sum_{S(t,t') \in {\cal C'}}
w_{S(t,t')}$ $=  \sum_{S(t,t') \in {\cal C}} w_{S(t,t')} = w$. So,
since the optimal solution minimizes $w$, $\Delta(D,D({\cal C}))$
is minimum and $D({\cal C})$ is an LS-fix.}

\defproof{Proposition \ref{prop:keep}}{From Propositions
A.\ref{prop:LFcombined} and A.\ref{prop:LScover}.}

\defproof{Proposition \ref{prop:allLSfix}}{To prove it it is
enough to construct this optimal cover. Let ${\cal C}=\{S(t,t')|
t' \in (D'\smallsetminus D)$. By definition ${\cal C}' = {\cal C}$
and $D({\cal C})=D'$. We need to prove that ${\cal C}$ is an
optimal cover. Since $D'$ is consistent, all the violation sets
were solved and therefore ${\cal C}$ is a cover. Also, since
$\Delta(D,D')=\Delta(D,D({\cal C}))= w$ and $\Delta(D,D')$ is
minimum, ${\cal C}$ minimizes the weight and therefore is an
optimal cover. }

\defproof{Proposition \ref{theo:ApproxNonCon}}{We have to
establish that the transformation of $\nit{DFOP}$ into
$\nit{MWSCP}$ given above is an $L$-reduction
\cite{papadimitriou94}. So, it remains to verify that the
reduction can be done in polynomial time in the size of instance
$D$ for $\nit{DFP}(\IC)$, i.e. that ${\cal G}$ can be computed in
polynomial time in $n$, the number of tuples in $D$. Notice that
if $m_i$ the number of database atoms in $\nit{ic}_i \in IC$, and
$m$ the maximum value of $m_i$ there are at most $n^{m_i}$
hyperedges associated to $\nit{ic}_i \in \IC$, each of them having
between $1$ to $m$ tuples.  We can check that the number of sets
$S(t,t')$ and their weights are polynomially bounded by the size
of $D$. There is one $S(t,t')$ for each local fix. Each tuple may
have no more than $|{\cal F}| \times |\IC|$ local fixes, where
${\cal F}$ is the set of fixable attributes.

The weight of each $S(t,t')$ is polynomially bounded by the
maximum absolute value in an attribute in the database and the
maximum absolute value of a constant appearing in $\IC$ (by an
argument similar to the one given in the proof of Proposition
\ref{lemma:squa-member}).

With respect to $D({\cal C})$, the number of sets in $\cal S$ is
polynomially bounded by the size of $D$, and since $\cal C
\subseteq S$, $\cal C$ is also polynomially bounded by
 the size of $D$. To generate $\cal C'$ it is necessary to search
 through $\cal S$. Finally, in order to replace $t$ in $D$ for each tuple
 $t'$ such that $S(t,t')\in {\cal C}$ we need to search through $D$.}

\vspace{.3cm}\defproof{Proposition \ref{prop:ApproxCover}}{Using
the same arguments as in the proof of Proposition
A.\ref{prop:LScover} we have that since $\hat{{\cal C}}$ is a
cover then $D({\hat{\cal C}})$ is a fix of $D$ wrt $\IC$. We need
to prove that $\Delta(D,D(\hat{\cal C})) \leq log(N) \times
\Delta(D,D')$. We know that  $\Delta(D,D({\cal \hat{C}})) =
\sum_{t \in D} \Delta(\{t\},\{t'\})$. $= \sum_{S(t,t') \in
\hat{\cal C}^\star} w_{S(t,t')}$. As described in definition
\ref{def:star}, $\hat{\cal C}^\star$ is obtained from $\hat{\cal
C}$ by replacing, for each $t$, all the sets $S(t,t_i) \in
\hat{\cal C}$ by a unique set $S(t,t^\star)$ such that
$S(t,t^\star)= \bigcup_{i} S(t,t_i)$. Since we are using euclidian
distance to calculate the local fixes, $\Delta(\{t\},\{t^\star\})
\leq  \sum_i \Delta(\{t\},\{t_i\})$. Then,

$\Delta(D,D(\hat{\cal C}))=\sum_{S(t,t') \in \hat{\cal C}^\star}
w_{S(t,t')}$ $\leq$ $\sum_{S(t,t') \in {\cal C}} w_{S(t,t')} =
\hat{w}$.

\noindent Thus, $\Delta(D,D(\hat{\cal C})) \leq \hat{w} \leq
log(N) \times w^o = log(N) \times \Delta(D,D')$, for every  LS-fix
$D'$ of $D$. }

\vspace{.3cm}\defproof{Theorem \ref{theo:forest}}{Based on the
tractability results in \cite{fuxman}, it suffices to show that
the LS-fixes for a database $D$ are in one-to-one and polynomial
time correspondence with the repairs using tuple deletions
\cite{ABC99,chomickiIC} for a database $D'$ wrt a set of key
dependencies.

Since we have 1ADs, the violation sets will have a single element,
then, for an inconsistent tuple $t$ wrt a constraint $\nit{ic} \in
\IC$, it holds ${\cal I}(D,\nit{ic},t)=\{t\}$. Since all the
violation sets are independent, in order to compute an LS-fix for
$D$, we have to generate independently all the local fixes $t'$
for all inconsistent tuples $t$ such that  $(\{t\},\nit{ic}) \in
S(t,t')$, with $\nit{ic} \in \IC$; and then combine them in all
possible ways.

Those local fixes can be found by considering all the {\em
candidate} fixes (not necessarily LS-minimal) that can obtained by
combining all the possible limits for each attribute provided by
the ICs (c.f. Proposition A.\ref{prop:prelimits}); and then
checking which of them satisfy $\IC$, and finally choosing those
that minimize $\Delta(\{t\},\{t'\})$. There are at most $2^{|{\cal
F}|}$ possible candidate fixes, where ${\cal F}$ is the set of
fixable attributes.

Let us now define a database $D'$ consisting of the consistent
tuples in $D$ together with all the local fixes of the
inconsistent tuples. By construction, $D$ and $D'$ share the same
keys.  Since each inconsistent tuple in $D$ may have more than one
local fix, $D'$ may become inconsistent wrt its key constraints.
Each repair for $D'$, obtained by tuple deletions, will choose one
local fix for each inconsistent tuple $t$ of $D$, and therefore
will determine an LS-fix of $D$ wrt $\IC$.}

\vspace{.3cm}\defproof{Proposition \ref{prop:possaggregsimple}}{
The $\nit{NP}$-complete $\nit{PARTITION}$ problem \cite{garey} can
be reduced to this case for a fixed set of 1ADs. Let a $A$ be a
finite set, whose elements $a$ have integer sizes $s(a)$. We need
to determine if there exists a subset $S$ of $A$, such that
$\sum_{a \in S} s(a) = n := (\sum_{a \in A} s(a))/2$.

We use two tables: $\nit{Set(Element,Weight)}$, with key
$\nit{\{Element,Weight\}}$, containing the tuples $(a,s(a))$ for
$a \in A$; and $\nit{Selection(Element}, X, Y)$, with key
$\nit{Element}$, fixable numerical attributes $X,Y$ (the partition
of $A$) taking values $0$ or $1$ (which can be specified with
1ADs), and initially containing the tuples $(a,0,0)$ for $a \in
A$. Finally, we have the  1AD $\forall E, X, Y \neg
(\nit{Selection}(E, X, Y), X < 1, Y < 1)$.

There is a one-to-one correspondence between LS-repairs of the
original database and partitions $X, Y$ of $A$ (collecting the
elements with value $1$ in either  $X$ or $Y$). Then, there is a
partition with the desired property iff the query $Q:~ (Set(E,W),
\nit{Selection}(E, X, Y), X = 1, \nit{sum}(W) = n)$~ has answer
${\it yes}$ under the brave semantics. The query used in this
proof is acyclic and belongs to the class ${\cal
C}_{\!\nit{Tree}}$.}

\vspace{.3cm}\noindent For the proof of Theorem \ref{theo:range}
we need some preliminaries. Let us define a function $F$, with
domain ${\cal G} \times S$, where ${\cal G} = \langle {\cal
V},{\cal E} \rangle$ is a graph and $S$ is a subset of vertexes of
graph $\cal G$, and range non-negative integers. The function is
defined as the summation over all the vertices $v \in S$, of cubes
of the number of edges connecting $v$ to vertexes in the
complement of $S$.
\begin{definition}\label{def:F} \em
Given a graph ${\cal G} = \langle {\cal V}, {\cal E} \rangle$ and
subset of its vertexes $S \in {\cal V}$
\begin{itemize}
    \item \vspace*{-.3cm} $F^l(S,v) = |T(S,v)|^3$ where $T(S,v)=$
    $\left \{
    \begin{array}{cc} \{v' |  v' \in  ({\cal V}
     \smallsetminus S)\land (v, v') \in {\cal E}), & ~v \in
     S \\
     \emptyset, & v \not \in S\end{array} \right.$
     \item $F({\cal G},S) =\sum_{v \in S} F^l(S,v)$\boxtheorem
\end{itemize}
\end{definition}

\begin{lemmaA}\label{lemma:Fbound}
Given a fixed regular undirected graph ${\cal G} = ({\cal V},
{\cal E})$ of degree $3$, the maximal value of $F({\cal G},S)$ on
all possible sets $S \subseteq V$ is $(3^3 \times |I|)$ for $I$ a
maximal independent set. \boxtheorem\end{lemmaA} \dproof{Lemma
A.\ref{lemma:Fbound}}{Let us first assume that $S$ is an
independent set, not necessarily maximal. In this case the
  answer to $F({\cal G}, S)$ will be $3^3 \times |S|$, because each element $v \in S$ is connected to
three vertices in ${\cal V} \smallsetminus S$. Then, among
independent sets, the maximum value for $F({\cal G}, S)$ is $3^3
\times m$, where $m$ is the maximum cardinality of an independent
set.

Let ${\cal G}[S]={\cal G}(S,{\cal E}_S)$ where ${\cal E}_S$ are
all the edges  $(v,v') \in {\cal E}$ such that $v,v' \in S$. Now,
if $S$ is not an independent set, there exists a maximum
independent set $I_S$ of ${\cal G}[S]$. Every $v \in
 ({\cal V} \smallsetminus S)$ is adjacent to at least one vertex in $I_S$, otherwise $I_S \cup \{v\}$,
 would be an independent set contained in $S$ and with more vertices  than $I_S$, contradicting our
choice of $I_S$. Now let us define $F_{ext}(S,v) = (F^l(S,v) +
\sum_{(v,v') \in {\cal E}}F^l(S,v'))$. Since every edge $v' \in (S
\smallsetminus I_S)$ is adjacent to $I_S$, it is easy to see that:
\begin{equation}\label{eq:Fext}F({\cal G},S) \leq \sum_{v \in I}
F_{ext}(S,v)\end{equation}

\noindent We want to prove that $F({\cal G},S) \leq F({\cal
G},I_S)$. This, combined with equation (\ref{eq:Fext}) shows that
it is enough to prove that $\sum_{v \in I_S} F_{ext}(S,v) \leq$ $
F({\cal G},I_S)$. Since $F({\cal G},I_S)= \sum_{v \in I_S}
F^l(I_S,v)$, we need to prove $\sum_{v \in I_S} F_{ext}(S,v) \leq
$ $\sum_{v \in I_S} F^l(I_S,v)$ and then,  it would be sufficient
to prove that $F_{ext}(S,v) \leq $ $F^l(I_S,v)$ is true for every
$v \in I_S$. For $v \in I_S$ and  $S{\,'} = (S \smallsetminus
 I_S)$, we have the following cases:
 \begin{enumerate}
  \item \label{it:1adj} If $v$ is adjacent to one vertex in $S{\,'}$ then
  $F_{ext}(S,v) \leq 2^3+2^3$ and $F^l(I_S,v)=3^3$ and therefore
  $F_{ext}(S,v) \leq (F^l(I_S,v)-11)$.
  \item If $v$ is adjacent to two vertexes in $S{\,'}$ in analogous way to item (\ref{it:1adj}) we
  get $F_{ext}(S,v) \leq (F^l(I_S,v)-10)$.
  \item If $v$ is adjacent to three vertexes in $S{\,'}$ in analogous way to item (\ref{it:1adj}) we
  get $F_{ext}(S,v) \leq (F^l(I_S,v)-3)$.
 \end{enumerate}

 Then, we have proved that $F_{ext}(S,v) \leq $ $F^l(I_S,v)$ and
 therefore that $F({\cal G},S) \leq F({\cal G},I_S)$. We also know
 that, since $I_S$ is an independent set, that $F({\cal G},S) \leq F({\cal G},I_S) \leq 3^3 \times
 m$. }\\

\vspace{.3cm}\defproof{Theorem \ref{theo:range}}{(a) For {\em
sum}:~ By reduction from a variation of {\em Independent Set}, for
graphs whose vertices have all the same degree. It remains ${\it
NP}$-hard as a special case of {\em Independence Set for Cubic
Planar Graphs} \cite{GareyJohnsonStockmayer}. Given an undirected
graph ${\cal G} = ({\cal V}, {\cal E})$ with degree $3$, and a
minimum bound $k$ for the size of the maximal independent set, we
create a relation ${\it Vertex}(V, C_1, C_2)$, where the key $V$
is a vertex and $C_1, C_2$ are fixable and may take values $0$ or
$1$, but are all equal to $0$ in the initial instance $D$. This
relation is subject to the denial ${\it IC}\!: \forall V, C_1, C_2
\neg({\it Vertex}(V,C_1,C_2), C_1 < 1, C_2 < 1)$. $D$ is
inconsistent wrt this constraint and in any of its LS-fix each
vertex $v$ will have associated a tuples ${\it Vertex}(v,1,0)$ or
${\it Vertex}(v,0,1)$ but not both.  Each LS-fix of the database
defines a partition of $\cal V$ into two subsets: $S$ with
$(v,1,0)$ and $S'$ with $(v,0,1)$, where clearly $S \cup S' =
{\cal V}$ and $S \cap S' = \emptyset$. Let us define a second
relation ${\it Edge}(V_1, V_2,W)$, with rigid attributes only,
that contains the tuples $(v_1, v_2, 1)$ for $(v_1,v_2) \in {\cal
E}$ or $(v_2,v_1) \in {\cal E}$. Every vertex $v$ appears in each
argument in exactly $3$ tuples.

Consider the ground aggregate conjunctive query $Q$:\\
\hspace*{1cm} $q({\it sum}(W_0)) \leftarrow  {\it  Vertex}(V_1, C_{11}, C_{12}),$ $C_{11} = 1,$ \\
 \hspace*{3.6cm}${\it Edge}(V_1, V_2,W_0),$  ${\it Vertex}(V_2, C_{21}, C_{22}),$  $ C_{21} = 0,$ \\
 \hspace*{3.6cm}${\it Edge}(V_1, V_3, W_1),$ ${\it Vertex}(V_3, C_{31}, C_{32}),$ $ C_{31} = 0,$ \\
 \hspace*{3.6cm}${\it Edge}(V_1, V_4, W_2), $ ${\it Vertex}(V_4, C_{41}, C_{42}),$ $ C_{41} = 0.$

 The query $Q$, computes the sum of cubes of the number of vertexes of $S'$ adjacent to vertices in
 $S$, i.e. it calculates the function from graph to nonnegative numbers
 corresponding to $F({\cal G},S)$ from Definition \ref{def:F} with $Q(D')=F({\cal
 G},S)$ for $D' \in Fix(D,\nit{IC})$ and $S=\{v | {\it Vertex}(v,1,0) \in
 D'\}$.

 We are interested in the minimum and maximum value for $Q$ in $Fix(D, {\it
 IC})$, i.e. the {\em min-max answer} introduced in \cite{ABC03}.
Since the function is nonnegative and since its value is zero for
$S=\emptyset$ and $S={\cal V}$ we have that its minimum value is
zero. We are only missing to find its maximum value.

From Lemma \ref{lemma:Fbound} we have that the answer to query Q
is at most $3^3 \times |I|$ with $I$ a maximum independent set. In
consequence, the min-max answer for $Q$ is (0,~ $3^3 \times m$),
with m the cardinality of the maximum independent set; and then
there is an independent set of size at least $k$ iff ~${\it
min\!\!-\!\!max~answer~to}~ Q \geq  k\times 3^3$.

\noindent (b) For {\em count distinct}:~ By reduction from {\em
MAXSAT}. Assume that an instance for {\em MAXSAT} is given,
consisting of a set $U$ of propositional variables, a collection
$C$ of clauses over $U$ and a positive integer $k$. The question
is whether at least $k$ clauses can be satisfied simultaneously,
which will get answer {\em yes} exactly when a question of the
form $\nit{countd} \leq (k-1)$, with $\nit{countd}$ defined by an
aggregate query over a database instance (both of them to be
constructed below), gets answer {\em no} under the {\em min-max}
semantics.

Define a relation $\nit{Var}(u, c_1, c_2)$, with (rigid) first key
attribute, and the second and third fixable (the denial below and
the minimality condition will make them take values 0 or 1). The
initial database has a tuple $(u, 0, 0)$ for every $u \in U$.
Another relation $\nit{Clause}(u, c, s)$, has no fixable
attributes and contains for every occurrence of variable $u \in U$
in a clause $c \in C$ a tuple $(u, c, s)$ with $s$ an assignment
for $u$ satisfying clause $c$. The IC is $\forall u, c_1, c_2 \neg
(\nit{Var}(u, c_1, c_2), c_1 < 1, c_2 < 1)$.  The acyclic query is
$$q(\nit{countd}(c)) \leftarrow
  \nit{Var}(u, c_1, c_2), \nit{Clause}(u, c, s), c_1 = s,$$
where $\nit{countd}$ denotes the ``count distinct" aggregate
function.  Its answer tells us how many clauses are satisfied in a
given LS-fix. The {\em max} value taken on a LS-fix, i.e. the {\em
min-max} answer, will be the {\em max} number of clauses which may
be satisfied for {\em MAXSAT}.

\noindent (c) For {\em average}:~ By reduction from $3SAT$. We use
the same table $\nit{Var}(u, c_1, c_2)$ and IC as in (a). Now, we
encode clauses as tuples in a fixed relation\\ $\nit{Clause(val,
var_1, val_1, var_2, val_2, var_3},$ $\nit{val_3)}$, where
$\nit{var_1, var_2, var_3}$ are the variables in the clause (in
any order), $\nit{val_1, val_2, val_3}$ all possible combinations
of truth assignments to variables (at most 8 combinations per
clause). And $\nit{val}$ is the corresponding truth value for the
clause ($0$ or $1$). Now, consider the acyclic query
\begin{eqnarray*}
 q(avg(val)) &\leftarrow& \nit{Clause(val, var_1, val_1, var_2, val_2, var_3, val_3)},\\
   && \nit{Var(var_1, val_1,val_1'), Var(var_2, val_2,val_2'), Var(var_3, val_3,val_3')}.
\end{eqnarray*}
Then value of $q$ is maximum in a LS-fix, taking value 1, i.e. the
{\em min-max} answer to $q$ is 1, iff the formula satisfiable.}\\

\defproof{of Theorem \ref{theo:approx-sum}}{
First we reduce {\em CQA} under range semantics for aggregate
queries with {\em sum} to $\nit{RWAE2}$, a restricted weighted
version of the problem of solving algebraic equations over
$\nit{GF}[2]$, the field with two elements. Next, we prove that
such an algebraic problem can be solved within constant
approximation factor. 

\noindent {\bf (A).}~ {\em Reduction to  $\nit{RWAE2}$}.
 In order
to define polynomial equations, we need variables. We introduce a
set $\cal V$ of variables $X^R_{k,i}$, taking values in
$\nit{GF}[2]$, for every tuple $t_i$ in an LS-fix corresponding to
a tuple $t$ (a ground database atom in the database) with key $k$
in a relation $R$ in the original database, i.e. $t_i$ belongs to
some LS-fix and $t_i$ $t$ share the key values $k$. For example if
the tuple $t$ is consistent or admits only one local fix (one
attribute can be changed and in only one way), only one variable
is introduced due to $t$. Denote with $\nit{bag}(t)$ the set of
variables introduced due to a same initial tuple $t$.

Consider a conjunctive query $$Q(sum(z)) :- R_1(\vec{x}), \cdots,
R_m(\vec{x}).$$ Throughout the a proof $\psi$ is the body of the
query  as a conjunction of atoms, $m$
  is the number of database predicates in $\psi$, $n$ is the number of
  tuples in the database, $k$ is the maximal number of attribute comparisons in the ICs
  (and the maximal number of fixes of a given tuple).

We may consider all the possible assignments $\beta$ from database
atoms in the query to grounds tuples in fixes that satisfy $\psi$.
The number of assignments is polynomial in the size of the
database,  actually $\leq n^m$. Notice that the the number of
LS-fixes of a database may be exponential, but the number of local
fixes of each original tuple is restricted by the number of
attributes of the tuple. So, the number of all possible LS-fixes
of tuples is polynomial in the size of the original database (even
linear). Here we are using the fact that we have 1ADs.

Now we build a system $\mathcal{E}$ of  weighted algebraic
  equations.
Each such assignment $\beta$ is associated with a combination of
tuples $t^{R_1}_{k_1,i_1}, \cdots, t^{R_m}_{k_m, i_m}$ satisfying
$\psi$. For each combination put the following equation $E^\beta$
over $\nit{GF}[2]$ into ${\cal E}$:
\begin{equation}\label{eq:eq}
 \overbrace{X^{R_1}_{k_1, i_1} \cdot \cdots X^{R_m}_{k_m, i_m}}^{selected} \cdot
 \underbrace{\prod_{i \not = i_1} (1 - X^{R_1}_{k_1, i})
 \cdot \cdots \cdot  \prod_{i \not = i_m} (1 - X^{R_m}_{k_m, i})}_{non-selected} ~=~
 1.
 \end{equation}
The first product in (\ref{eq:eq}), before the first $\prod$,
contains the variables corresponding to the tuples selected by
$\beta$. The rest of the product contains variables for the those
tuples that were not selected, i.e. if $t_1$ appears in the first
product, with $t_1 \in bag(t)$, and $t_2 \in bag(t)$, with $t_1
\neq t_2$, then the variable $X_2$ corresponding to $t_2$ appears
as  $(1 - X_2)$ in the second part of the product. This captures
the restriction that no two different tuples from the same bag can
be used (because the share the key values). For each combination
$\beta$ of tuples in LS-fixes there is no more then one equation,
which in turn has a polynomial number of factors.

Equation (\ref{eq:eq}) gets weight $w(E^\beta)$ that is equal to
the value of aggregation attribute $z$ in $\beta$.

In this way we have an instance of the $\nit{RWAE2}$. It requires
to find the maximum weight for a subsystem of $\cal E$ that can be
(simultaneously) satisfied in $\nit{GF}[2]$, where the weight of
the subsystem is the sum of the weights of the individual
equations. Of course, this problem also has a version as a
decision problem, so as {\em CQA} under range semantics.

\noindent {\bf Claim:}~ The maximal weight of a satisfied
subsystem of $\cal E$ is the same as the maximal value of
$Q(\nit{sum}{z})$ over all possible LS-fixes of $D$.

\noindent ($\geq$)~ Assume that query $Q$ takes a maximum value
over all possible LS-fixes of  $D$ on
 an  LS-fix $D'$. Under 1ADs a database LS-fix $D'$ is a set union of local fixes, with
 one local fix selected for every original tuple. Consider an assignment $A$
defined on $\mathcal{V}$ that maps
  variables corresponding a selected local fix to $1$ and all other variables to $0$.

 Consider all sets of local fixes which simultaneously satisfy  $\psi$. If local fixes $t_1, \cdots, t_m$
 satisfy $\psi$, then there exist exactly one equation $e$ for that given set of
local fixes. The equation $e$ will be
 satisfied since variables corresponding to selected local fixes have value $1$, and
 ``non-selected"
 variables have value $0$. So, for every set of local fixes satisfying the query body, there would be
 a satisfied equation with  weight equal to the value of aggregated attribute. It means, that a solution
 to the algebraic equation problem is bigger or equal to the maximal query answer ({\em min-max} answer).

 \noindent ($\leq$)~ Consider an assignment $A$ which is a solution of algebraic equation problem. It maps
 elements of $\mathcal{V}$ to $\{0,1\}$ in such a way that the weight of satisfied equations of $\mathcal{E}$ is
 maximum over all possible assignment for $\mathcal{V}$.

 First we prove that if there exists a bag $B$ such that more then one of its variables is
 mapped to $1$, then there exist an assignment $A'$ with the same weight of satisfied equations
 of $\mathcal{E}$ as $A$, but $B$ contains no more then one variable mapped to $1$.

 Assume that for a bag $B$ more then two variables (let us say $X_i,
 X_j$) are mapped to $1$. It means that every equation which contains
 variables from $B$ will be unsatisfied, since it contains either
 $(1-X_i)$ or $(1-X_j)$ as factors in the equation. If we change a value of
 one of the variables (say $X_i$) to $0$, then no satisfied equation
 become unsatisfied, since satisfied equations do not contain $X_i$.
 No unsatisfied equation becomes satisfied, because due to the
 assumption of
 maximality of the weight of the satisfied subset of $E$ for $A$.

 In a second step, we prove that if $A$ is a maximal assignment and there exist a bag $B$
 such that all of its variables are mapped to $0$, then there exist an assignment $A'$,
 which satisfies the same subset of $\mathcal{E}$ as $A$, but at least one variable
 from that $B$ is mapped to $1$.

 If all variables from a bag $B$ are mapped to $0$, then all equations
 which contain variables from $B$ are unsatisfied. If we change a
 value of one variable to $1$, then no satisfied equation becomes
 unsatisfied since all satisfied equations do not contain variables
 from $B$. No unsatisfied equation becomes satisfied due to maximality
 assumption of the weight of satisfied equation for $A$.
 Taking step by step all bags from $\mathcal{V}$, for given a maximum assignment $A$, we
 produce an assignment $A'$, which has exactly one variable from each bag mapped to 1.

Now, construct a database $D'$ which is a set of local fixes
corresponding to variables mapped to $1$.
 It is obviously a LS-fix, and  $w(E(A)) \leq Q(D')$.

\noindent {\bf (B).}~  \textit{A deterministic approximation
algorithm for $\nit{RWAE2}$.} ~The construction and approximation
factor obtained are similar those in the approximation of {\em
MAXSAT}.
 C.f. \cite{vazirani,papadimitriou94}. In two steps, first a randomized
 algorithm is produced, that is next de-randomized.

 \noindent {\bf (B1).}~ \textit{Randomized approximation
 algorithm}.~
 Assume that from each bag we select one variable with probability $1/k$, where $k$ is the
 number of variables in the bag. We map selected variable to $1$ and all other variables in the bag to $0$.
 For each equation $e$, random variable $W_e$ denotes the weight contributed by $e$ to the total
 weight $W$. Thus, $W = \sum_{e \in \mathcal{E}} W_e$ and $\mathbf{E}[W_e] = w_e
 \cdot \mathbf{Pr}[\textit{e\ is\ satisfied}]$, where $\mathbf{E}$ is a mathematical
 expectation and $\mathbf{Pr}$ is a probability.

 If the query contains $m$ predicates, then each equation contains no more than $m$ variables
 from different bags (never two different variables from the same bag), then $\mathbf{E}(W_e) \geq k^{-m} w_e$.
 Now, by linearity of expectation,
\[
\mathbf{E}[W] = \sum_{e \in \mathcal{E}}\mathbf{E}[W_e] \geq
k^{-m} \sum_{e \in \mathcal{E}} w_e \geq k^{-m} \cdot OPT.
\]
\noindent {\bf (B2).}~ \textit{De-randomization via conditional
expectation.}~ We first establish

\noindent {\bf Claim:}~ The \textit{RWAE2} problem is
self-reducible \cite[Chap. A.5]{vazirani}.

In fact, assume $A'$ is a partial assignment from $\mathcal{V}$,
 such that variables $X_1, \cdots, X_i$ are mapped to $\{0, 1\}$. Let ${\cal E}^s$ be the set of equations
 satisfied by $A'$ with  total weight $W[E^s]$, and ${\cal E}^u$ is the set of equations which cannot
 be satisfied under $A'$. Let $E''$ be a set of equations from ${\cal E} \smallsetminus ({\cal E}^s \cup {\cal E}^u)$,
 such that variables from $X_1, \cdots, X_i$ are replaced by their
values. By additivity of the weight
 function and the independence of the variables, the maximal weight of satisfied equations under an assignment
 which extends $A'$ is $W[E^s] + max W[E'']$, where $W[E'']$ is a solution of the \textit{RWAE2} problem
 restricted to $E''$. It is good enough to consider the self-reducibility trees $T$
such only one variable from each
 bag gets value $1$ along any path in the tree. This establishes our claim.

Assume that a self-reducibility tree $T$ is given, with each node
in it corresponding to a step of the self-reduction. Each node $v$
of $T$ is labelled with $X_1 = a_1, \cdots, X_i = a_i$, a partial
assignment of values to variables $X_1, \cdots, X_i \in
\mathcal{V}$ associated to the step for $v$ of the self-reduction.
Since this is a partial assignment, some of the equations in $\cal
E$ become immediately satisfied, other unsatisfied, and some other
undetermined. The latter become a set of equations $E'$ associated
to $v$ on variables ${\cal V} \smallsetminus \{X_1, \ldots,
X_i\}$, obtained from $\cal E$ by giving to the variables $X_1,
\ldots, X_i$ their values $a_1, \ldots, a_i$. By construction,
these equations inherit the weight of the corresponding equations
in $\cal E$.

For example, if the set of equations consists of: (1),
$yp(1-x)=1$, (2) $2 xz(1-y) = 1$, (3) $ 3 xw(1-y) = 1$, with
variables $x, y, z, p, w$, and the partial assignment, at some
step of self-reduction for $v$ is  $x = 1, y = 0, w = 1$, then
equation (1) becomes unsatisfiable, (2) is not satisfied but
possibly satisfiable with an appropriate value for $z$; and (3)
satisfied. So, $E'$ contains equation (2), but with $x,y$ replaced
by their values $1,0$, resp.

The conditional expectation of any node $v$ in $T$ can be computed
via its sets of equations $E'$ we just described. Clearly, the
expected weight of satisfied equations of $E'$ under a random
assignment of values in $\nit{GF}[2]$ to $\mathcal{V}\setminus
X_1, \cdots, X_i$ can be computed in the polynomial time. Adding
to this the weight of the equations in $\cal E$ already satisfied
by the partial assignment $X_1 = a_1, \cdots, X_i = a_i$ gives the
conditional expectation.

Then we compute in polynomial time a path from the root to a leaf,
such that the conditional expectation of each node on this path is
$\geq \mathbf{E}[W]$. This can be done as in the  construction in
\cite[Theorem 16.4]{vazirani}.

In consequence, we can find a deterministic  approximate solution
to the \textit{RWAE2} problem in polynomial time.  It approximates
the optimum solution with a factor greater then $k^{-m}$. It means
that we can approximate the maximal value of aggregate conjunctive
query within a factor $k^{-m}$, which depends on integrity
constraints and a query, but not depend on the size of the
database. This ends the proof.

For example, the query with {\em sum} used in the proof of the
{\em NP}-hardness in Theorem \ref{theo:range} has $m = 4, k = 2$,
then it can be approximated  within the factor $2^{-4}$.}

\subsection{An Example for Theorem \ref{theo:exist}}\label{sec:hilbert}

Consider the diophantine equation
\begin{equation}
2x^3y^2 + 3xy + 105 = x^2y^3+y^2. \label{eq:dioph} 
\end{equation}
Each term $t$ in it will be represented by a relation $R(t)$ with
8 attributes taking values in $\mathbb{N}$: three, $X_1, X_2,
X_3$, for the maximum exponent of $x$, three, $Y_1, Y_2, Y_3$, for
the maximum exponent of $y$, one, $C$, for the constant terms,
plus a last one, $K$, for a key. Value 0 for a non-key attribute
indicates that the term appears in $t$, otherwise it gets value 1.
We introduce as many tuples in $R(t)$ as the coefficient of the
term; they differ only in the key value. We will see that only the
0 values will be subject to fixes. These are the relations and
their ICs:

\begin{center}
\begin{tabular}{c|cccccccc}
$R(2x^3y^2)$&$X_1$&$X_2$&$X_3$&$Y_1$&$Y_2$&$Y_3$&$C$&$K$ \\

&0 & 0 & 0 & 1 & 0 & 0& 1&1\\
&0 & 0 & 0 & 1 & 0 & 0& 1&2\\
\end{tabular}
\end{center}
For this table we have the following set, $\IC(2x^3y^2)$, of
ICs:\\
$\forall x_1 \cdots x_8  \neg(R(2x^3y^2)(x_1, \ldots, x_8) \land
x_1 \neq x_2)$, $\forall x_1 \cdots x_8  \neg(R(2x^3y^2)(x_1,
\ldots, x_8) \land x_2 \neq x_3)$,\\ $\forall x_1 \cdots x_8
\neg(R(2x^3y^2)(x_1, \ldots, x_8) \land x_5 \neq x_6)$, $\forall
x_1 \cdots x_8  \neg(R(2x^3y^2)(x_1, \ldots, x_8) \land x_4 \neq
1)$,\\ $\forall x_1 \cdots x_{16}  \neg(R(2x^3y^2)(x_1, \ldots,
x_8) \land R(2x^3y^2)(x_9,
\cdots, x_{16}) \land x_1 \not = x_9)$ \\
$\forall x_1 \cdots x_{16}  \neg(R(2x^3y^2)(x_1, \ldots, x_8)
\land R(2x^3y^2)(x_9, \cdots, x_{16}) \land x_5 \not = x_{13}).$

\begin{center}
\begin{tabular}{c|c c c c c c c c}
$R(3xy)$ & $X_1$  & $X_2$ & $X_3$ & $Y_1$ & $Y_2$ & $Y_3$ & $C$ & $K$ \\
& 1 & 1 & 0 & 1 & 1& 0 & 1& 3\\
& 1 & 1 & 0 & 1 & 1& 0 & 1& 4\\
& 1 & 1 & 0 & 1 & 1& 0 & 1& 5\\
\end{tabular}
\end{center}

\noindent $\IC(3xy)$:\\
$\forall x_1 \cdots x_{16}  \neg(R(3xy)(x_1, \ldots, x_8) \land
R(3xy)(x_9, \ldots,
x_{16}) \land x_3 \not = x_{11})$, \\
$\forall x_1 \cdots x_{16}  \neg(R(3xy)(x_1, \ldots, x_8) \land
R(3xy)(x_9, \cdots,
x_{16}) \land x_6 \not = x_{14})$,\\
$\forall x_1 \cdots x_{8}  \neg(R(3xy)(x_1, \ldots, x_8) \land x_1
\not = 1) $,
$\forall x_1 \cdots x_{8}  \neg(R(3xy)(x_1, \ldots, x_8) \land x_2 \not = 1)$, \\
$\forall x_1 \cdots x_{8}  \neg(R(3xy)(x_1, \ldots, x_8) \land x_4 \not = 1)$, \\
$\forall x_1 \cdots x_{8}  \neg(R(3xy)(x_1, \ldots, x_8) \land x_5
\not = 1).$

\begin{center}
\begin{tabular}{c |c c c c c c c c}
$R(105)$ & $X_1$  & $X_2$ & $X_3$ & $Y_1$ & $Y_2$ & $Y_3$ & $C$ & $K$ \\
& 1 & 1 & 1 & 1 & 1& 1& 105 & 6\\
\end{tabular}
\end{center}

\noindent$\IC(105)$:

$\forall x_1 \cdots x_{8}  \neg(R(105)(x_1, \ldots, x_8) \land x_1
\not = 1)$, $\forall x_1 \cdots x_{8}  \neg(R(105)(x_1, \ldots,
x_8) \land x_2 \not = 1)$,\\
$\forall x_1 \cdots x_{8}  \neg(R(105)(x_1, \ldots, x_8) \land x_3
\not = 1)$,  $\forall x_1 \cdots x_{8}  \neg(R(105)(x_1, \ldots,
x_8) \land x_4 \not = 1)$,\\
$\forall x_1 \cdots x_{8}  \neg(R(105)(x_1, \ldots, x_8) \land x_5
\not = 1)$,  $\forall x_1 \cdots x_{8}  \neg(R(105)(x_1, \ldots,
x_8) \land x_6 \not = 1)$,\\  $\forall x_1 \cdots x_{6}
\neg(105(x_1, \cdots, x_6) \land x_7 \not = 105)$.

Similar tables $R(x^2y^3)$ and $R(y^2)$ and corresponding sets of
ICs are generated for the terms on the RHS of (\ref{eq:dioph}).

Next we need ICs that are responsible for making equal all $x$s
and $y$s in
all terms of the equation:\\
$\forall x_1 \cdots x_{16}  \neg(R(2x^3y^2)(x_1, \ldots, x_8)
\land
R(3xy)(x_9, \cdots, x_{16})  \land x_1 \not = x_{11})$,\\
$ \forall x_1 \cdots x_{16}  \neg(R(2x^3y^2)(x_1, \ldots, x_8)
\land
 R(3xy)(x_9, \ldots, x_{16})  \land x_5 \not = x_{13})$\\
$ \forall x_1 \cdots x_{16}  \neg(R(2x^3y^2)(x_1, \ldots, x_8)
\land
 R(x^2y^3)(x_9, \cdots, x_{16})  \land x_1 \not = x_{10})$\\
$ \forall x_1 \cdots x_{16}  \neg(R(2x^3y^2)(x_1, \ldots, x_8)
\land
 R(x^2y^3)(x_9, \ldots, x_{16})  \land x_5 \not = x_{12})$\\
$ \forall x_1 \cdots x_{16}  \neg(R(2x^3y^2)(x_1, \ldots, x_8)
\land
 R(y^2)(x_9, \ldots, x_{16})  \land x_5 \not = x_{13}).$

Now we construct a single table $R(\nit{equ})$ that represents
equation (\ref{eq:dioph}) by appending the previous tables:

\begin{center}
\begin{tabular}{c |c c c c c c c c}
$R(\nit{equ})$ & $X_1$  & $X_2$ & $X_3$ & $Y_1$ & $Y_2$ & $Y_3$ & $C$ & $K$ \\
&0 & 0 & 0 & 1 & 0 & 0& 1&1\\
&0 & 0 & 0 & 1 & 0 & 0& 1&2\\
& 1 & 1 & 0 & 1 & 1& 0 & 1& 3\\
& 1 & 1 & 0 & 1 & 1& 0 & 1& 4\\
& 1 & 1 & 0 & 1 & 1& 0 & 1& 5\\
& 1 & 1 & 1 & 1 & 1& 1& 105 & 6\\
& 1 & 0 & 0 & 0 & 0& 0& 1 & 7\\
& 1 & 1 & 1 & 1 & 0& 0& 1 & 8
\end{tabular}
\end{center}
We need ICs stating the correspondence between the terms in the
tables $R(t)$ and table $R(\nit{equ})$:\\
$ \forall x_1 \cdots x_{16}  \neg(R(\nit{equ})(x_1, \ldots, x_8)
\land
 R(2x^3y^2)(x_9, \ldots, x_{16}) \land x_8 = x_{16} \land x_1 \not =
 x_9)$,\\
$ \forall x_1 \cdots x_{16}  \neg(R(\nit{equ})(x_1, \ldots, x_8)
\land
 R(2x^3y^2)(x_9, \ldots, x_{16}) \land x_8 = x_{16} \land x_2 \not = x_{10})$,\\
$\cdots ~~~~~~~~~~~ \cdots ~~~~~~~~~~~~ \cdots$\\
$ \forall x_1 \cdots x_{16}  \neg(R(\nit{equ})(x_1, \ldots, x_6)
\land
 R(y^2)(x_7 \cdots x_{16})\land ~x_8 = x_{\!16} \land x_{7} \not =
 x_{15}).$

Finally, we have one aggregate constraint that is responsible for
making equal the LHS and RHS of equation (\ref{eq:dioph}):

\noindent $sum_{R(\nit{equ})}(x_1\cdot x_2\cdot x_3 \cdot x_4
\cdot x_5 \cdot x_6 \cdot x_7~:x_6 < 7) ~=
~~~~sum_{R(\nit{equ})}(x_1\cdot x_2 \cdot x_3 \cdot x_4 \cdot x_5
\cdot x_6 \cdot x_7~:x_6>6).$

If the database has an LS-fix, then there is an integer solution
to the diophantine equation. If the equation has a solution $s$,
then there is an instance $R(equ)^\prime$ corresponding to $s$
that satisfies the ICs. By Proposition \ref{lemma:existence},
there is an LS-fix of the database.

The reduction could be done with the table $R(\nit{equ})$ alone,
making all the ICs above to refer to this table, but the
presentation would be harder to follow.

\end{document}